\documentclass[prd,preprint,superscriptaddress,amsmath,amssymb]{revtex4}
\pdfoutput=1

\usepackage{graphicx,color,slashed}% Include figure files
\begin{document}

%{\begin{flushright}{KIAS-P17017}
%\end{flushright}}

\title{\bf \Large  Study of  $B^\pm_c \to  (D^0 K^\pm,  D^0 \pi^\pm)$ decays }

\author{Chuan-Hung Chen}
\email{physchen@mail.ncku.edu.tw}
\affiliation{Department of Physics, National Cheng-Kung University, Tainan 70101, Taiwan}

\author{Yen-Hsun Lin}
\email{chrislevel@gmail.com}
\affiliation{Department of Physics, National Cheng-Kung University, Tainan 70101, Taiwan}

\date{\today}% It is always \today, today,

\begin{abstract}
LHCb observes the $B^+_c \to D^0 K^+$ decay with $R_{D^0 K} = f_c/f_u \times {\cal B}(B^+_c \to D^0 K^+)=(9.3^{+2.8}_{-2.5} \pm 0.6)\times 10^{-7}$. The  corresponding branching ratio (BR)  of the decay  can be estimated as ${\cal B}(B^+_c \to D^0  K^+) \approx (10.01 \pm 3.40)\times 10^{-5}$; however, the theoretical estimates vary from $\sim 10^{-7}$ to $\sim 5\times 10^{-5}$. We phenomenologically investigate the $B^+_c \to (D^0 K^+, D^0 \pi^+)$ decays through the analysis of $B\to KK$, $B^+_u\to D^+  K^0$, and $B_d \to D^-_s K^+$. With the form factor of $f^{B_c D}_0\approx 0.2$, it is found that the tree-annihilation contribution dominates the $B^+_c \to D^0 K^+$ decay, and when ${\cal B}(B^+_u \to D^+  K^0) \approx (1-3.1) \times 10^{-7}$ is required, we obtain ${\cal B}(B^+_c \to D^0 K^+)\approx (4.4- 9) \times 10^{-5}$, and the magnitude of CP asymmetry is lower than approximately $10\%$.  Although the $B^+_c \to D^0 \pi^+$ decay is dominated by the tree-transition effect, the tree-annihilation also makes an important contribution, where its effect could be around $70\%$ of the tree-transition. It is found that when ${\cal B}(B^+_c \to D^0 K^+)\approx (4.4- 9) \times 10^{-5}$ is taken, the BR and CP asymmetry for $B^+_c \to D^0 \pi^+$ with the common values of parameters can be ${\cal B}(B^+_c \to D^0 \pi^+)\approx (4.9-8)\times 10^{-6}$ and of the order of one, respectively. Moreover, we conclude ${\cal B}(B^+_c \to D^+ K^0)\approx {\cal B}(B^+_c \to D^0 K^+)$,  and the BRs for $B^+_c \to K^+  \bar K^0$ and $B^+_c \to J/\Psi \pi^+$ are $(6.99 \pm 1.34) \times 10^{-7}$ and $(7.7 \pm 1.1)\times 10^{-4}$, respectively. 
\end{abstract}

\maketitle

\section{Introduction}

With a data sample of  $3.0$ fb$^{-1}$ at $\sqrt{s}=7$ and 8 TeV, LHCb recently  observes the $B^+_c \to D^0 K^+$ decay, and the observable with a statistical significance of $5.1\sigma$ is given as~\cite{Aaij:2017kea}:
\begin{equation}
R_{D^0 K}=\frac{f_c}{f_u} {\cal B}(B^+_c \to D^0 K^+) =(9.3^{+2.8}_{-2.5} \pm 0.6)\times 10^{-7}\,, \label{eq:LHCb}
\end{equation}
where $f_{c(u)}$ denotes the transition probability  of a $b$-quark hadronizing to a $B_{c(u)}$, and ${\cal B}(B^+_c \to D^0 K^+)$ is the branching ratio (BR) of $B^+_c \to D^0 K^+$. The involved Cabibbo-Kobayashi-Maskawa (CKM) matrix elements  are $V^*_{us} V_{ub}$ and $V^*_{cs} V_{cb}$ for the tree interactions and $V^*_{ts} V_{tb}$ for the loop penguin interactions. Based on the measurement in Eq.~(\ref{eq:LHCb}), the ratio of the branching fraction of $B^+_c \to D^0 K^+$ to $B^+_c \to J/\psi \pi^+$ is obtained as $R_{DK/J/\psi\pi}={\cal B}(B^+_c \to D^0 K^+)/{\cal B}(B_c \to J/\psi \pi^+)= 0.13\pm 0.04 \pm 0.01 \pm 0.01$, where the third error is from the $R_{J/\psi \pi^+}$ measurement. 
 
If we use  ${\cal B}(B^+_c \to J/\Psi \pi^+) \approx 7.7\times 10^{-4}$ ( see  later analysis) as an input, the current LHCb measurement indicates that  ${\cal B}(B^+_c \to D^0 K^+)$ is in the region of $(6.74-13.3)\times 10^{-5}$ when  a $1\sigma$ error of $R_{DK/J/\psi \pi}$ is taken.  However,  the  theoretical estimations  are quite uncertain  even in terms of the order of magnitude; for instance,  ${\cal B}(B^+_c \to D^0 K^+)\sim 5\times 10^{-5}$ was achieved by~\cite{Du:1998te,Zhang:2009ur}; $\sim (0.3, 2) \times 10^{-7}$ were obtained by~\cite{Choi:2009ym,Fu:2011tn}, and  $\sim 5 \times 10^{-6}$ was estimated by~\cite{Rui:2011qc}. Although \cite{Du:1998te} and \cite{Zhang:2009ur} can predict the results of $O(10^{-5})$, the origin used to obtain the large BR is different; the former relies on the loop penguin with a large $B^+_c \to D^0$ form factor, $f^{B_c D}_0(m^2_K) \sim 0.60$, in the QCD factorization approach, and the latter relies on the tree-annihilation process in the perturbative QCD, in which the resulting $B^+_c \to D^0$ form factor is $f^{B_c D}_0(m^2_K) \sim 0.22$. From~\cite{Zhang:2009ur}, the tree-annihilation may play a main role in the $B^+_c \to D^0 K^+$ decay. 

 In view of the very different results of ${\cal B}( B^+_c \to D^0 K^+)$ in the literature, in this work, we investigate the $B^+_c \to D^0 K^+$ decay using a phenomenological approach. General discussions with flavor symmetry can be found in~\cite{Bhattacharya:2017aao}. In terms of  flavor diagrams, it is found that with the exception of CKM matrix elements,  the  $B^+_c \to D^0 K^+$ decay can be classified by four different topological flavor diagrams, which include: tree-transition $(T_T)$, tree-annihilation $(A^{c}_T)$, penguin-transition $(T^{u}_P)$, and penguin-annihilation $(E^{c}_P)$. The contribution of each diagram  to the decay amplitude can be decomposed into factorizable and nonfactorizable parts. In order to understand the influence of each flavor diagram and each (non)factorization piece, we search for the measured processes in $B$ decays, for which the associated flavor diagrams are similar to those in $B^+_c \to D^0 K^+$. Based on these measured $B$ decays, we can analyze the relative sizes of the  topological diagrams, and using the associated Wilson coefficients (WCs) and color suppression factor,  small sub-leading  effects can be dropped, and only dominant contributions are retained. We then apply the obtained results to the $B^+_c \to D^0 K^+$ decay. Since the decay amplitudes from the tree- and penguin-transition are usually dominated by the factorizable parts, which are clearer in theoretical calculations, we will focus on the contributions from the annihilation topologies.

  We find that the $B\to KK$, $B^+_u \to D K^+$, and $B_d \to D^-_s K^+$ decays can be the potential candidates in our analysis. Some interesting properties of $B\to KK$ can be revealed in a phenomenological analysis, and they are summarized as: (i)  when a penguin-annihilation is neglected due to the small WCs and color suppression, $B_d \to K^+ K^-$ is dominated by the nonfactorizable tree-annihilation flavor diagram, and via a proper parametrization, this nonfactorization effect can be applied to the $B^+_u \to K^+ \bar K^0$ decay; (ii) the same  approximation in (i), ${\cal B}(B_d \to  K^0 \bar K^0)$,  which arises from the pure penguin contributions, can be completely determined by the experimental data; (iii) ${\cal B}(B^+_u \to K^+ \bar K^0)$  can be formulated just in terms of the BRs of $B_d \to K^+ K^-$ and $B_d \to  K^0 \bar K^0$, and if we neglect the small ${\cal B}(B_d \to K^+ K^-)$, we obtain ${\cal B}(B^+_u \to K^+ \bar K^0) \approx \tau_{B_u}/\tau_{B_d} {\cal B}(B_d \to K^0 \bar K^0)$, which fits well with the experimental data; (iv) although we can not predict the strong phase, using the phenomenological analysis,  the CP asymmetry (CPA) of $B^+_u \to K^+ \bar K^0$, which is induced by the interference between the tree-annihilation and penguin effect, can be estimated to be $|A_{CP}(B^+_u \to K^+ \bar K^0)| \lesssim 10\%$. Hence, from the analysis of the $B\to KK$ decays, we can clearly see how large the nonfactorizable tree-annihilation contribution can be.  

Although the $B^+_u \to D^0 K^+$ and $B_d \to D^-_s K^+$ decays are pure tree-annihilation processes,  both topological flavor diagrams and the associated WCs are different. The flavor diagram of $B^+_u \to D^0 K^+$ is similar to the tree-annihilation in $B^+_u \to K^+ \bar K^0$ while $B_d \to D^-_s K^+$ is close to $B_d \to K^+ K^-$. That is, when we take the  characteristic effects of  charmed mesons into account, the parametrization for the nonfactorization effect used in $B\to KK$ can be applied to  the $D^0 K^+$ and $D^-_s K^+$ modes. Due to $m_D \gg m_K$,  unlike the situation in $B\to KK$, the factorizable tree-annihilation contributions may not be negligible. Since the WC of the factorizable part in $B_d \to D^-_s K^+$  is smaller than that in $B^+ \to D^0 K^+$, it is found that when we drop  the factorization effect in $B_d \to D^-_s K^+$, we obtain ${\cal B}(B_d \to D_s K^-)/{\cal B}(B_d \to K^+ K^-)\sim 285$, which coincides with the experimental data.  On the contrary, the BR of $B^+_u \to D^0 K^+$ strongly depends on the factorization effect, e.g., ${\cal B}(B^+_u \to D^0 K^+)$ can be of $O(10^{-7})$ and $O(10^{-9})$ with and without the factorization contribution, respectively. Although $B^+_u \to D^0 K^+$ has not yet been observed, we can use the extracted result, which is based on the upper limit of $B^+_u \to D^0 K^{*+}$, to bound the  free parameters in our analysis. 

 When  the nonfactorizable  tree-annihilation contribution to $DK$ modes is determined from the $B_d \to (K^+ K^-, D_s K^-)$ decays, and   the factorizable part is bounded from the $B^+_u \to D^0 K^+$ decay, we can then estimate the BR and CPA for the  $B^+_c\to D^0 K^+$ decay. We find: (a)  the tree-annihilation topological diagram dominates the others, where ${\cal B}(B_c \to D^0 K^-)$ can be $\sim 6 \times 10^{-5}$ and  $\sim 10^{-5}$ with and without the tree-annihilation contribution, respectively; (b) the factorizable tree-annihilation is larger than its nonfactorizable contribution; (c) ${\cal B}(B^+_c \to D^0 K^+)$ can be as large as $9 \times 10^{-5}$ when ${\cal B}(B^+_u \to D^0 K^+) < 3.1 \times 10^{-7}$ is satisfied; (d) the CPA of $B^+_c \to D^0 K^+$ can be less than around $10\%$, where the value depends on ${\cal B}(B^+_c \to D^0 K^+)$ and ${\cal B}(B^+_u \to D^0 K^+)$. Moreover, we  apply the same approach to $B^+_c \to D^+ K^0$ and $B^+_c \to D^0 \pi^+$, where we obtain ${\cal B}(B^+_c \to D^+ K^0)\approx {\cal B}(B^+_c \to D^0 K^+)$, ${\cal B}(B^+_c \to D^0 \pi^+) = (4.9-8) \times 10^{-6}$, and the magnitude of CPA for $B^+_c \to D^0 \pi^+$ can be $O(1)$. 

The paper is organized as follows: In Sec.~II, we phenomenologically study the $B\to KK$ decays. The time-like form factors from vector and scalar currents for the annihilation processes are defined. We also parametrize and determine the nonfactorizable parts of the annihilation flavor diagrams for the $B_d \to K^+ K^-$ and $B^+_u \to K^+ \bar K^0$ decays. In Sec.~III, we study the $B^+_u \to D^0 K^+$ and $B_d \to D^-_s K^-$ decays, which include the influence of factorizable tree-annihilation. In Sec.~IV, we analyze the $B^+_c \to D^0 K^+$ decay in detail. The relative magnitudes of various topological flavor diagrams are presented. The applications to the decays $B^+_c \to D^+ K^0$, $B^+_c \to K^+  \bar K^0$, $B^+_c \to J/\psi \pi^+$, and $B^+_c \to D^0 \pi^+$ are discussed. A summary is given in Sec.~V.

\section{ Phenomenological analysis of the $B\to KK$ decays}

Hereafter, we will use anti-$B$-meson decays to present our analysis;  thus, the quark contents of $\bar B_d$ and $B^-_u$  are $b\bar d$ and $b\bar u$, respectively, unless stated otherwise. The current measurements of BRs for the $B\to KK$ decays are~\cite{PDG}:
 \begin{align}
 {\cal B}(\bar B_d \to \bar K^0  K^0)^{\rm exp} & = (1.21 \pm 0.16)\times 10^{-6} \,,  \nonumber \\
 {\cal B}(B^-_u \to K^- K^0)^{\rm exp} & = (1.31 \pm 0.17)\times 10^{-6} \,, \nonumber \\
{\cal B}(\bar B_d \to K^- K^+)^{\rm exp} & = (7.8 \pm 1.5)\times 10^{-8} \,. \label{eq:dataBKK}
 \end{align}
It can be seen that the difference in BR between the $\bar K^0  K^0$ and $K^- K^0$ modes is only around $8\%$. We will demonstrate that this difference mainly arises from the  lifetimes of $B_u$ and $B_d$ when the tree annihilation effect in $B^-_u \to K^- K^0$ is neglected  due to small factors, such as the CKM matrix element $V_{ub}$ and the effective WC  $C_1/N_c$. Such a topological annihilation diagram will also contribute to the $B^-_c \to \bar D^0 K^-$ process; however,  this tree annihilation effect on the $B_c$ decay becomes crucial when the CKM matrix element $V_{cb}$ and factorizable tree-annihilation are properly taken into account. 

The effective Hamiltonian for  the $B\to KK$ decays, which is from the $W$-mediated tree and  the gluonic penguin  diagrams, is written as~\cite{Buchalla:1995vs}:
 \begin{equation}
 {\cal H} = \frac{G_F}{\sqrt{2}}V^*_{ud} V_{ub} \left( C_1(\mu)O_1 + C_2(\mu) O_2 \right) -\frac{G_F}{\sqrt{2}} V^*_{td} V_{tb} \sum^6_{i=3} C_i(\mu) \sum_q O^q_i \,.\label{eq:HBKK}
  \end{equation}
$V_{ij}$ are the CKM matrix elements, and the values used in the paper are shown in Table~\ref{tab:WCs}.  $C_j(\mu)$ are the WCs at $\mu$ scale, and  their values at $\mu=2.5$ GeV with a naive dimensional regularization (NDR) scheme are shown in Table.~\ref{tab:WCs}~\cite{Buchalla:1995vs,Ali:1997nh}. The operators are given as:
  \begin{align}
  O_1 & = (\bar d_\beta u_\alpha)_{V-A} (\bar u_\alpha b_\beta)_{V-A}\,, \ O_2 = (\bar d_\beta u_\beta)_{V-A} (\bar u_\alpha b_\alpha)_{V-A}\,, \nonumber \\
O^q_3 & = (\bar q_\beta q_\beta)_{V-A} (\bar d_\alpha b_\alpha)_{V-A}\,, \  O^q_4 =  (\bar q_\beta q_\alpha)_{V-A} (\bar d_\alpha b_\beta)_{V-A}\,, \nonumber \\
O^q_5 & =  (\bar q_\beta q_\beta)_{V+A} (\bar d_\alpha b_\alpha)_{V-A}\,, \  O^q_6 = (\bar q_\beta q_\alpha)_{V+A} (\bar d_\alpha b_\beta)_{V-A}\,,
  \end{align}
where $(\bar f' f)_{V\pm A}=\bar f' \gamma_\mu (1\pm \gamma_5) f$, and  $\alpha(\beta)$ are the color indices.  Since the electroweak penguin effects in these decays are small, we neglect their contributions in the analysis. A detailed discussion with complete operators can be found in~\cite{Chen:2000ih}.

\begin{table}[htp]
\caption{Values of CKM matrix elements used in the study, where $\gamma=70^{\circ}$ and $\beta=22^{\circ}$ are taken. Wilson coefficients (WCs) at $\mu=2.5$ GeV for $m_{t,\rm pole}=175$ GeV with NDR scheme~\cite{Ali:1997nh}.}
\begin{tabular}{ccccccc} \hline\hline
CKM & $V_{ud,cs,tb}$ & $V_{us(cd)}$ & $V_{ub}$  & $V_{cb}$ & $V_{td}$ & $V_{ts}$ \\ \hline 
    & $\approx 1$  & $0.22(-0.22)$~ & $0.0037 e^{-i\gamma} $ & 0.04 & $ 0.0084 e^{-i\beta}$ & $-0.04$ \\ \hline \hline
WC & $C_1$ & $C_2$ & $C_3$ & $C_4$ & $C_5$  & $C_6$ \\ \hline
    & ~~$-0.257$~~ & ~~1.117~~ & ~~0.017~~  & ~~$-0.044$~~ & ~~0.011~~ & ~~$-0.056$~~ \\ \hline \hline

\end{tabular}
\label{tab:WCs}
\end{table}%

\begin{figure}[phtb]
\includegraphics[scale=0.7]{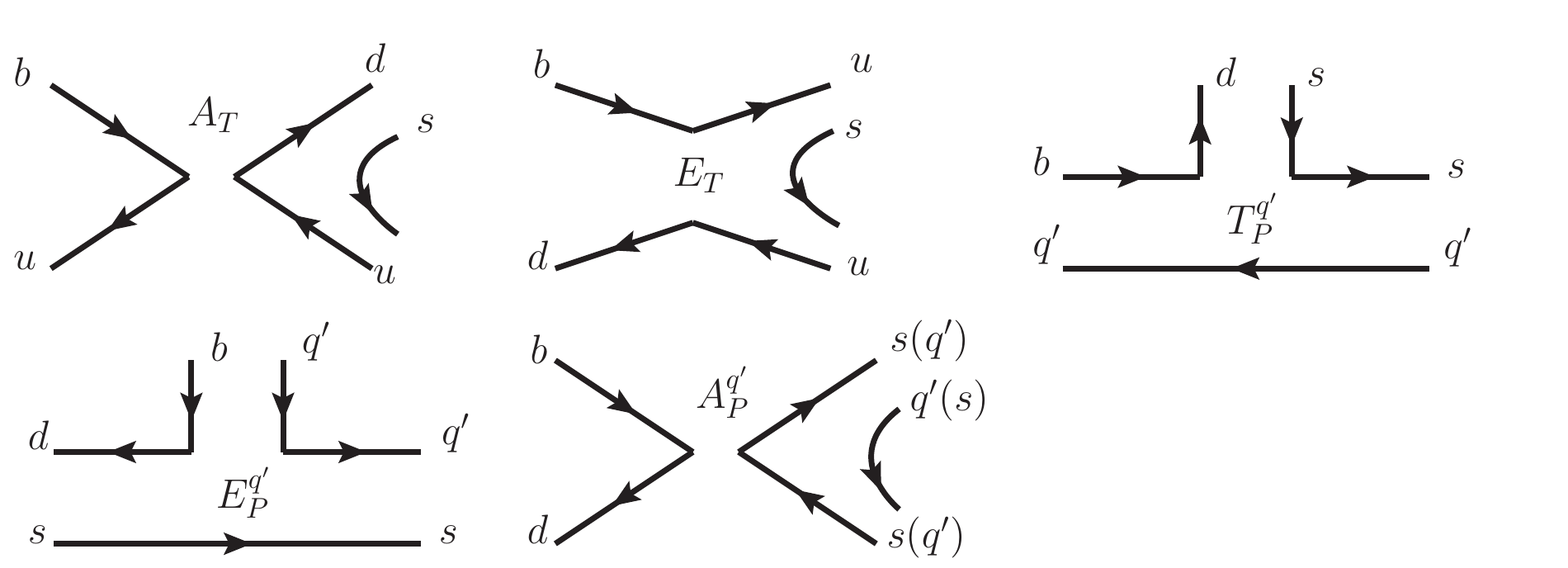}
 \caption{Flavor diagrams for the $B\to KK$ decays with $q'=u,d$.  }
\label{fig:TDBKK}
\end{figure}

According to the effective interactions in Eq.~(\ref{eq:HBKK}), we find that the $B\to KK$ decays can be classified into five types of topological flavor diagrams, and  the  diagrams are shown in Fig.~\ref{fig:TDBKK}, where $A^{(q')}_{T(P)}$ and $E^{(q')}_{T(P)}$ ($q'=u,d$) denote the annihilation topologies from the tree (T)  and penguin (P) contributions, respectively, and $T^{q'}_P$ represents the contributions from the penguin-transition flavor diagram.  Thus, the decay amplitudes for $B\to KK$ can be written as:
 \begin{align}
 M(\bar K^0  K^0) &= - \frac{G_F}{\sqrt{2} }V^*_{td} V_{tb} \left( T^d_P + E^d_P + A^d_P\right)\,, \nonumber \\
  M(K^-  K^0) &= \frac{G_F}{\sqrt{2} }V^*_{ud} V_{ub} A_T - \frac{G_F}{\sqrt{2} }V^*_{td} V_{tb} \left( T^u_P + E^u_P \right)\,, \nonumber \\
   M(K^-  K^+) &= \frac{G_F}{\sqrt{2} }V^*_{ud} V_{ub} E_T -\frac{G_F}{\sqrt{2} } V^*_{td} V_{tb}  A^u_P\,. \label{eq:MBKK}
 \end{align}
  Each component in a decay amplitude can be decomposed into factorizable and nonfactorizable parts. Since   the associated WCs in these parts are different,  for clarity, we show their relations in Table~\ref{tab:F-NF-WCs}, where  $a_{1,2}=C_{2(1)}+ C_{1(2)}/N_c$, $a_{4(6)}=C_{4(6)}+ C_{3(5)}/N_c$, $a_{3(5)}=C_{3(5)}+ C_{4(6)}/N_c$, and $N_c=3$ is the number of colors. 
   
   %%%%

\begin{table}[htp]
\caption{The associated Wilson coefficients of the factorizable part (FP) and nonfactorizable part (NFP) in each topological diagram, where $a_{1,2}=C_{2(1)}+ C_{1(2)}/N_c$; $a_{4(6)}=C_{4(6)}+ C_{3(5)}/N_c$;  $a_{3(5)}=C_{3(5)}+ C_{4(6)}/N_c$, and $N_c=3$ is the number of colors. }
\begin{tabular}{c|cccccccc} \hline\hline
DA & $T^d_P$ &  $E^d_P$ & $A^d_P$ & $A_T$ & $T^u_P$  & $E^u_P$ & $E_T$ & $A^u_P$ \\ \hline
 FP    &  $a_{4,6}$ & $a_{4,6}$ & $a_{3,5}$ & $a_1$  & $a_{4,6}$ & $a_{4,6}$ & $a_2$ & $a_{3,5}$\\ \hline 
 NFP  &  ~$C_{3,5}/N_c$~ & ~$C_{3,5}/N_c$~ & ~$C_{4,6}/N_c$~ & ~$C_1/N_c$~ & ~$C_{3,5}/N_c$~ & ~$C_{3,5}/N_c$~ & ~$C_1/N_c$~ & ~$C_{4,6}/N_c$~ \\ \hline \hline

\end{tabular}
\label{tab:F-NF-WCs}
\end{table}%

In order to discuss  the relations among the decay amplitudes shown in Eq.~(\ref{eq:MBKK}), we parametrize the time-like form factors for two pseudoscalar mesons in the final state as:
 \begin{equation}
 \langle P_1(p_1) P_2(p_2)| \bar q_2 \gamma_\mu q_1 | 0 \rangle  = F_1(Q^2) \left( q_\mu - \frac{Q\cdot q}{Q^2} Q_\mu  \right) + F_0(Q^2) \frac{Q\cdot q}{Q^2} Q_\mu\,, \label{eq:tlFF}
 \end{equation}
where $Q=p_1+p_2$, $q=p_2-p_1$, and $F_{1,0}(Q^2)$ are the time-like form factors.  As a result, we obtain $Q^\mu \langle P_1 P_2 | \bar q_2 \gamma_\mu q_1 |0\rangle = (m^2_2-m^2_1) F_0(Q^2)$. When the $P_2$ meson is the CP-conjugated state  of $P_1$, we get $Q^\mu \langle P \bar P | \bar q \gamma_\mu q |0\rangle =0$.   Based on this result, it can be concluded that the factorizable part of the annihilation topology  induced by $(V-A) \otimes (V-A)$ is suppressed by $m^2_2-m^2_1$.   According to Eq.~(\ref{eq:tlFF}),    the time-like form factor of a scalar current can be parametrized as:
 \begin{equation}
 \langle P_1(p_1) P_2(p_2)| \bar q_2 \, q_1| 0 \rangle  = \frac{Q\cdot q}{m_{q_2} - m_{q_1} } F_0(Q^2)\,. \label{eq:tlFFS}
 \end{equation}
 Clearly, although there is a suppression factor $m^2_2- m^2_1$  in the numerator,  an enhancement factor $1/(m_{q_2} -m_{q_1})$ for the light quarks appears; thus, Eq.~(\ref{eq:tlFFS}) could be sizable. Since the scalar current can be generated from  the Fierz transformation of  $(V+A) \otimes (V-A)$,  the factorizable part of annihilation topology  induced from $(V+A) \otimes (V-A)$  may not be suppressed. 
 
 According to Eqs.~(\ref{eq:tlFF}) and (\ref{eq:tlFFS}), we  now discuss the $A^{u(d)}_P$ and $E^{u(d)}_P$ effects. Since the behavior of $A^u_P (E^d_P)$ is the same as that of $A^d_P(E^u_P)$,  we only focus on $A^d_P$ and $E^u_P$ in the following analysis. The operators $O_3-O_6$ contributing to $A^d_P$ are derived through the vector currents; from the result of Eq.~(\ref{eq:tlFF}), the associated factorizable parts vanish. Therefore, $A^{d}_P$ has only nonfactorizable part and is given as:
  \begin{align}
  A^{d}_P &=\langle K^0 \bar K^0 | \sum^6_{k=3} C_k (O^s_k + O^d_k)  | \bar B_d \rangle \nonumber \\ 
% & \approx (a_3 + a_5)  \langle K^0 \bar K^0 | (\bar s \gamma^\mu s + \bar d \gamma^\mu d) |0\rangle (-i f_B p_{B\mu}) 
%\nonumber \\
 & \approx \langle K^0 \bar K^0 | \sum_{m=4,6}\frac{C_m}{N_c}  (O^s_m + O^d_m)  | \bar B_d \rangle_{NF}\,.
  \end{align}
$A^d_P$ in general is not zero; however, comparing it to the $T^{q'}_P$ effect, which is related to $C_{4,6}$, the $A^d_P$  contributions are suppressed by $C_{4,6}/N_c$. Since  no other possible enhancement factor appears, we assume that $A^{d, u}_P$ are negligible in the $B\to KK$ decays.  From Table~\ref{tab:F-NF-WCs}, the nonfactorizable part of $E^u_P$ is associated with $C_{3,5}/N_c$. Due to $|C_{3(5)}|< |C_{4(6)}|$, we also take $(E^u_P)_{NF} \approx 0$. To estimate the factorizable part of $E^u_P$,  the operators $O_3-O_6$ in $E^u_P$ have to  make the Fierz transformations.  The current-current interaction structures of $O_{3,4}$ are still $(V-A)\otimes (V-A)$ after the Fierz transformations; according to earlier discussions, their contributions are  thus suppressed by $m^2_{K^0} - m^2_{K^+}$ and can be neglected. In contrast to $O_{3,4}$, the $O_{5,6}$ operators become $(S+P)\otimes (S-P)$ when  the Fierz transformations are applied;  hence, their contributions  are sizable and can be expressed as:
 \begin{align}
 E^u_P &= \langle K^- K^0 | \sum^6_{k=3} C_k O^u_k | B^-_u \rangle  \nonumber \\
% & \approx - 2 a_6 \langle K^-  K^0 | \bar d (1+ \gamma_5) u |0\rangle \langle 0 |\bar u (1-\gamma_5) | B_u \rangle\nonumber \\
 % &+ \langle K^-  K^0 | \sum_{m=3,5}\frac{C_m}{N_c}  O^u_m   | B_u \rangle_{NF} \nonumber \\
  &  \approx 2 i a_6  \frac{f_B m^2_B}{m_b +m_u} \frac{m^2_{K^0} -m^2_{K^+}}{m_d -m_u}  F^{KK}_0(m^2_B)\,. \label{eq:EuP}
  %+ \langle K^-  K^0 | \sum_{m=3,5}\frac{C_m}{N_c}  O^u_m   | B_u \rangle_{NF} \,.
 \end{align}
 With $m_{K^0}=0.498$ GeV, $m_{K^\pm}=0.494$ GeV, and $m_{d(u)}=10(5)$ MeV, the factor $m^2_{K^0} -m^2_{K^\pm}/(m_d -m_u) \approx 0.79$ is not suppressed. 
 
 If we drop the $A^{d,u}_P$ contributions,  it can be seen from Eq.~(\ref{eq:MBKK}) that $\bar B_d \to K^- K^+$ is a tree-annihilation  process ($E_T$).  The tree-annihilation effect $A_T$ causes the difference between the $\bar K^0  K^0$ and $K^- K^0$ modes at the amplitude level. Since the similar topological diagrams $A_T$ and $E_T$ will respectively contribute to the  $B^-_c \to \bar D^0 K^-$  and $\bar B_d \to D^+_s K^-$ decays with the exception of the CKM matrix elements,  it is of interest to understand the relative size between $A_T$ and $E_T$ in $B\to KK$.  The interaction structures in $A_T$ and $E_T$  are $(V-A)\otimes (V-A)$; therefore,  the factorizable parts in both topologies are either suppressed or  vanished. Hence, $A_T$ and $E_T$ are dominated by the nonfactorizable parts. From Table~\ref{tab:F-NF-WCs} and $|C_{1}| < C_2$, we can obtain $A_T/E_T \sim C_1/C_2 \sim -0.23$.  With isospin symmetry, it can be expected that $T^d_P \approx T^u_P$ and $E^d_P\approx E^u_P$. If we set $T^d_P + E^d_P = T^u_P + E^u_P \equiv M_{TE}$ and take it as  a free parameter, using the data  in Eq.~(\ref{eq:dataBKK}) and the approximation of $A_T/E_T \sim C_1/C_2$, we can determine  $A_T$, $M_{TE}$, and the strong phase to be:
  \begin{align}
\frac{|V^*_{ud} V_{ub}A_T|}{|V^*_{td} V_{tb} M_{TE}|} & =  \left| \frac{C_1}{C_2} \right| \sqrt{\frac{{\cal B}(\bar B_d \to K^+ K^-)}{{\cal B}(\bar B_d \to K^0 \bar K^0)}} \sim 0.058\,,  \nonumber \\
{\cal B}^{\rm avg}_{ K^- K^0} & = \frac{\tau_{B_u}}{\tau_{B_d}} \frac{C^2_1}{C^2_2} {\cal B}_{K^- K^+} + \frac{\tau_{B_u}}{\tau_{B_d}} {\cal B}_{\bar K^0  K^0} + 2 \frac{\tau_{B_u}}{\tau_{B_d}}  \frac{C_1}{C_2} \cos(\delta)\cos(\alpha) \sqrt{{\cal B}_{K^- K^+}  {\cal B}_{\bar K^0  K^0}}\,, \label{eq:CRBKK}
  \end{align}
where ${\cal B}_{f}$ is the BR for the $B\to f$ decay; ${\cal B}^{\rm avg}_{ K^- K^0} =({\cal B}_{K^- K^0} +{\cal B}_{K^+ \bar K^0})/2$, $\alpha+\beta+\gamma=\pi$ is used, and $\delta$ is the relative strong phase of $A_T$ and $M_{TE}$. In addition, the CP asymmetry (CPA) of $K^- K^0$ mode can be expressed as:
 \begin{align}
 A_{CP}(K^- K^0) & = \frac{ {\cal B}_{K^- K^0} - {\cal B}_{K^+ \bar K^0 }}{{\cal B}_{K^- K^0} + {\cal B}_{K^+ \bar K^0 } }
 \nonumber \\
 &  = 2 \frac{\tau_{B_u}}{\tau_{B_d}}  \frac{C_1}{C_2} \sin(\delta)\sin(\alpha) \frac{ \sqrt{{\cal B}_{K^- K^+}  {\cal B}_{\bar K^0 K^0 }}}{{\cal B}^{\rm avg}_{K^- K^0}}\,.
 \end{align}

\begin{figure}[phtb]
\includegraphics[scale=0.5]{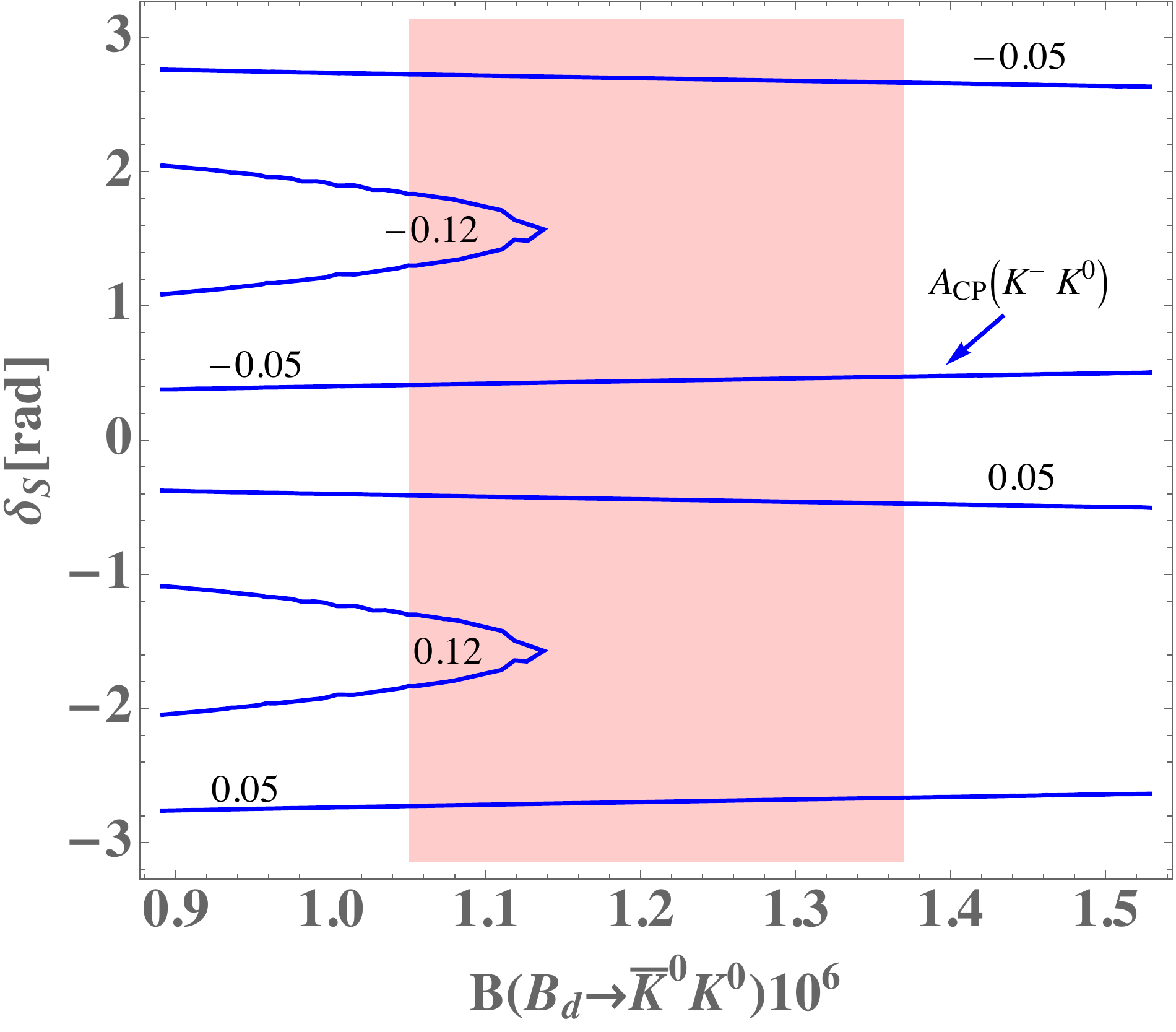}
 \caption{ Contours for $A_{CP}(K^0 K^-)$ (solid) as a function of ${\cal B}(K^0 \bar K^0)$ and $\delta$, where $\alpha=88^{\circ}$ is used, and the vertical band is ${\cal B}^{\rm exp}_{K^0 \bar K^0}$ with  $1\sigma$ errors. }
\label{fig:ACPKK}
\end{figure}

 From Eq.~(\ref{eq:CRBKK}),  it is known that the  $B^-_u\to K^- K^0$ decay has a strong correlation to the decays $\bar B_d\to (\bar K^0 K^0, K^- K^+)$. Due to ${\cal B}_{K^- K^+}\ll {\cal B}_{\bar K^0 K^0}$, we can drop the first term in Eq.~(\ref{eq:CRBKK}). Since $\alpha=(88\pm 5)^{\circ}$~\cite{Aaij:2015dwl} is close to $90^{\circ}$ and $|\cos(\alpha)|\leq 0.12$,  the third term in Eq.~(\ref{eq:CRBKK}) should be at most $2\%$ of $\tau_{u}/\tau_{d}{\cal B}_{K^0 \bar K^0}$. If we also neglect this term, Eq.~(\ref{eq:CRBKK}) becomes ${\cal B}_{K^- K^0} \approx \tau_{B_u}/\tau_{B_d} {\cal B}_{\bar K^0 \bar K} = 1.078 {\cal B}_{\bar K^0  K^0}$, where $\tau_{B_u, B_d}=(1.638, 1.52)$ ps are used, and the result fits very well with the current experimental measurements.  To numerically show the CPA of $K^0 K^-$ mode, we can take ${\cal B}_{K^- K^+}$, ${\cal B}_{\bar K^0 K^0}$, and $\delta$ as free parameters. Due to ${\cal B}_{K^- K^+}\ll {\cal B}_{\bar K^0 K^0}$, we fix ${\cal B}_{K^- K^+}=7.8 \times 10^{-8}$.  Thus, the contours for $A_{CP}(K^- K^0)$ (solid) as a function of ${\cal B}_{\bar K^0 K^0}$ and $\delta$ are shown in Fig.~\ref{fig:ACPKK}, where $\alpha=88^{\circ}$ is used, and the vertical band denotes ${\cal B}^{\rm exp}_{\bar K^0 K^0}$ with  $1\sigma$ errors. We can not determine $\delta$ well; therefore, the CPA can be in the range $|A_{CP}| \lesssim 12\%$. The result is consistent with  the current experimental value of $A_{CP}(K^- K^0)=-0.087 \pm 0.100$, averaged by the heavy flavor averaging group (HFLAV)~\cite{Amhis:2016xyh}. 

%%%%

\section{ Branching ratios for  $B^-_u \to  D^- \bar K^0$ and $\bar B_d \to D_s K^-$ }

Based on the study of the $B\to KK$ decays, we find some  characteristics of  annihilation topological diagrams in a $B$-meson decaying into two light pseudoscalars; that is,  the contribution from topology $E_T (E_P)$ is more significant than that from $A_T (A_P)$.  
%%We will apply these properties to the $B_c \to \bar D^0 K^-$ decay and understand the magnitude of BR. 
It is of interest to investigate if the property is preserved in the $B^-_c \to \bar D^0 K^-$ decay. 
Before we study $B^-_c \to \bar D^0 K^-$,  we investigate  the  $B$-meson decay processes, in which a charmed-meson and a $K$-meson are involved   in the final state,  and only annihilation topologies dictate the contributions.  We find that  the $B_u\to D^- \bar K^0$ and $B_d \to D_s K^-$ decays match our requirements, where their current experimental measurements are:
 \begin{align}
 {\cal B}(B^-_u \to D^- \bar K^0 )^{\rm exp} & < 2.9 \times 10^{-6}  \ \text{(PDG~\cite{PDG})}\,, \nonumber \\
 {\cal B}(\bar B_d \to D_s K^-) ^{\rm exp}& = (2.21 \pm 0.25)\times 10^{-5}\ \text{(HFLAV~\cite{Amhis:2016xyh})}\,.
  \end{align}
 We note that although the upper bound of  ${\cal B}(B^-_u \to D^- \bar K^0 )$ is of the order of $10^{-6}$,  LHCb reported ${\cal B}(B^-_u \to D^- K^{*0}) < 4.9 \times 10^{-7}$~\cite{Aaij:2015dwl}.  Since the BRs of the annihilation processes $\bar B_d \to D^{(*)}_s K^{(*)-}$  are close to each other, it is reasonable to conjecture that the upper limit of the BR for $B^-_u \to D^{-} \bar K^0$ could be:
 \begin{equation}
 {\cal B}(B^-_u \to D^- \bar K^0) \sim \frac{{\cal B}(\bar B_d \to D_s  K^-)}{{\cal B}(\bar B_d \to D_s  K^{*-})}  {\cal B}( B^-_u\to D^- \bar K^{*0}) < 3.1 \times 10^{-7}\,,
  \end{equation}
 where ${\cal B}(\bar B_d \to D_s  K^-)=2.21\times 10^{-5}$ and ${\cal B}(\bar B_d \to  D_s K^{*-})=3.50\times 10^{-5}$~\cite{PDG}  are used.
 
%\subsection{$B_u \to  D^- \bar K^0$ and $B_d \to D_s K^-$}

\begin{figure}[phtb]
\includegraphics[scale=1]{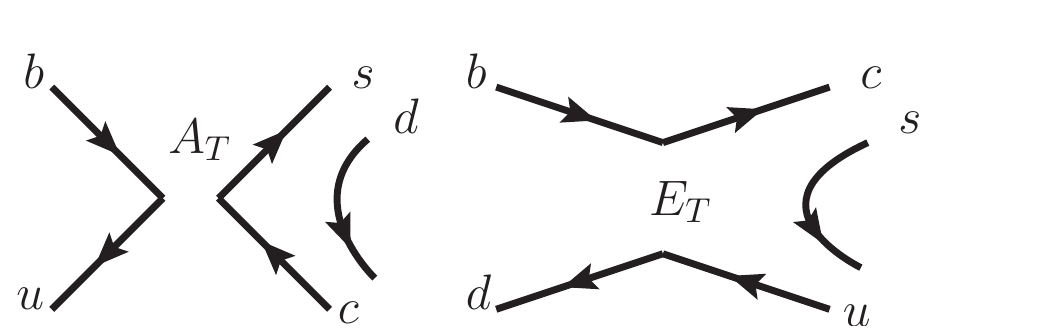}
 \caption{ Flavor diagrams for the $B^-_u \to D^- \bar K^0$ and $\bar B_d \to  D_s K^-$ decays. }
\label{fig:Topo_BDK}
\end{figure}

 The effective interactions for $B^-_u \to  D^- \bar K^0$ and $\bar B_d \to D_s K^-$ can be written as:
  \begin{align}
  {\cal H}_{DK} & = \frac{G_F}{\sqrt{2}}  V^*_{cs} V_{ub} \left( C_1(\mu) Q_1 + C_2 (\mu) Q_2\right)\nonumber \\
  &+ \frac{G_F}{\sqrt{2}}  V^*_{du} V_{cb} \left( C_1(\mu) Q'_1 + C_2 (\mu) Q'_2\right)\,, \label{eq:HDK}
  \end{align}
 where the effective operators are:
 \begin{align}
  Q_1 & = (\bar s_\beta c_\alpha)_{V-A} (\bar u_\alpha b_\beta)_{V-A}\,, \ Q_2 = (\bar s_\beta c_\beta)_{V-A} (\bar u_\alpha b_\alpha)_{V-A}\,, \nonumber \\
Q'_1 & = (\bar d_\beta u_\beta)_{V-A} (\bar c_\alpha b_\alpha)_{V-A}\,, \  Q'_2 =  (\bar d_\beta u_\alpha)_{V-A} (\bar c_\alpha b_\beta)_{V-A}\,.
  \end{align}
In terms of the flavor diagrams, which are shown in Fig.~\ref{fig:Topo_BDK}, it can be seen that $B^-_u \to  D^- \bar K^0$ and $\bar B_d \to D_s K^-$  arise from $A_T$ and $E_T$, respectively. Thus, the decay amplitudes can be parametrized as:
\begin{equation}
M(D^- \bar K^0)  = \frac{G_F}{\sqrt{2}}V^*_{cs} V_{ub} A^{DK}_T\,, \  M(D_s K^-)  =  \frac{G_F}{\sqrt{2}}V^*_{ud} V_{cb} E^{D_s K}_T\,. \label{eq:DKMM}
\end{equation}

 %%%%

 According to earlier discussions,  the factorizable parts of both decays indeed are proportional to $(m^2_K-m^2_{D_{(s)}})/m^2_{B}$. Due to $m_{D_{(s)}} \gg m_K$, it may not be a good approximation to directly drop these effects. However, by comparing the associated WCs, it can be seen that the associated WCs in the $D^- \bar K^0$ and $D_s K^-$ modes are $a_1$ and $a_2$, respectively.  Due to $a_2\ll a_1$, the factorizable part of $E^{D_s K}_T$ can be neglected as a leading approximation. For the nonfactorizable parts,   taking the similar assumption of $A_T/E_T\sim C_1/C_2$ used in $B\to KK$,  we use $(A^{DK}_T)_{NF}/(E^{D_s K}_T)_{NF} \sim f_D C_1/(f_{D_s} C_2)$, where in order to show the properties of  the $D$ and $D_s$ mesons, we include the decay constants of the charmed mesons with $f_D=0.209$ GeV and $f_{Ds} \approx 0.248$ GeV~\cite{PDG}. In order to explicitly describe the factorizable and nonfactorizable parts of $A^{DK}_T$, we must further parametrize these hadronic effects. With the time-like form factors defined in Eq.~(\ref{eq:tlFF}), we write $A^{DK}_T$ and $E^{D_s K}_T$ as:
 \begin{align}
  A^{DK}_T & = \langle D^- \bar K^0 | \sum_i C_i Q_i | B_u\rangle_{F+NF} \nonumber \\
  & \approx -i f_B \left[ a_1  (m^2_D -m^2_K) F^{DK}_{0}(m^2_B) - \frac{ C_1}{ N_c}  q^2 \chi_{D K}(m^2_B)\right] \,,  \label{eq:ADKT}\\
   E^{D_s K}_T  & \approx  (E^{D_s K}_T)_{NF} =  \langle D_s K^- | \sum_i C_i Q'_i | B_d\rangle_{NF} \nonumber \\
  & \approx  \frac{C_2}{N_c} \left[ i f_B  q^2  \chi_{D_s K}(m^2_{B}) \right] \,, \label{eq:EDsKT}
 \end{align}
where the form factors $\chi_{DK}$ and  $\chi_{D_s K}$ are from the nonfactorizable effects and are defined as:
  \begin{align}
   \langle D^-  \bar K^0 |  (\bar s_\alpha c_\beta)_{V-A}  (\bar u_\beta b_\alpha)_{V-A}|B^-_u\rangle_{NF} = i f_B \left[ Q\cdot  q\, \chi'_{DK}(Q^2) + q^2\, \chi_{DK}(Q^2)  \right]\,, \nonumber \\
    \langle D_s K^- |  (\bar c_\alpha u_\beta)_{V-A} (\bar d_\beta b_\alpha)_{V-A} |\bar B_d\rangle_{NF} = i f_B \left[ Q\cdot q \, \chi'_{D_s  K}(Q^2) + q^2 \, \chi_{D_s K} (Q^2) \right]\,, \label{eq:TLNF}
  \end{align}
 where $Q\cdot q = m^2_K-m^2_{D_{(s)}} $, and $q^2=2(m^2_{D_{(s)}} + m^2_K) -m^2_B$. Due to $|Q\cdot q| \ll |q^2|$, we exclude the $\chi'_{D_{(s)}K}$ contributions. The assumption of $(A^{DK}_T)_{NF}/(E^{D_s K})_{NF} \sim f_D C_1/(f_{D_s} C_2)$ leads to $\chi_{DK}(m^2_B)/\chi_{D_s K}(m^2_B) \sim f_{D}/f_{D_s}$.  Using Eq.~(\ref{eq:EDsKT}) and the values in Table~\ref{tab:WCs}, the magnitude of $\chi_{D_s K}(m^2_B)$ can be determined from ${\cal B}(\bar B_d \to D_s K^-)^{\rm exp}$ as: 
  \begin{equation}
  |\chi_{D_s K}(m^2_B)| = 0.119 \pm 0.010\,,
  \end{equation}
 where the error is  from the uncertainty  of ${\cal B}(\bar B_d \to D_s K^-)^{\rm exp}$. The strong phase of $\chi_{D_s K}$ cannot be directly determined in this approach.

\begin{figure}[phtb]
\includegraphics[scale=0.5]{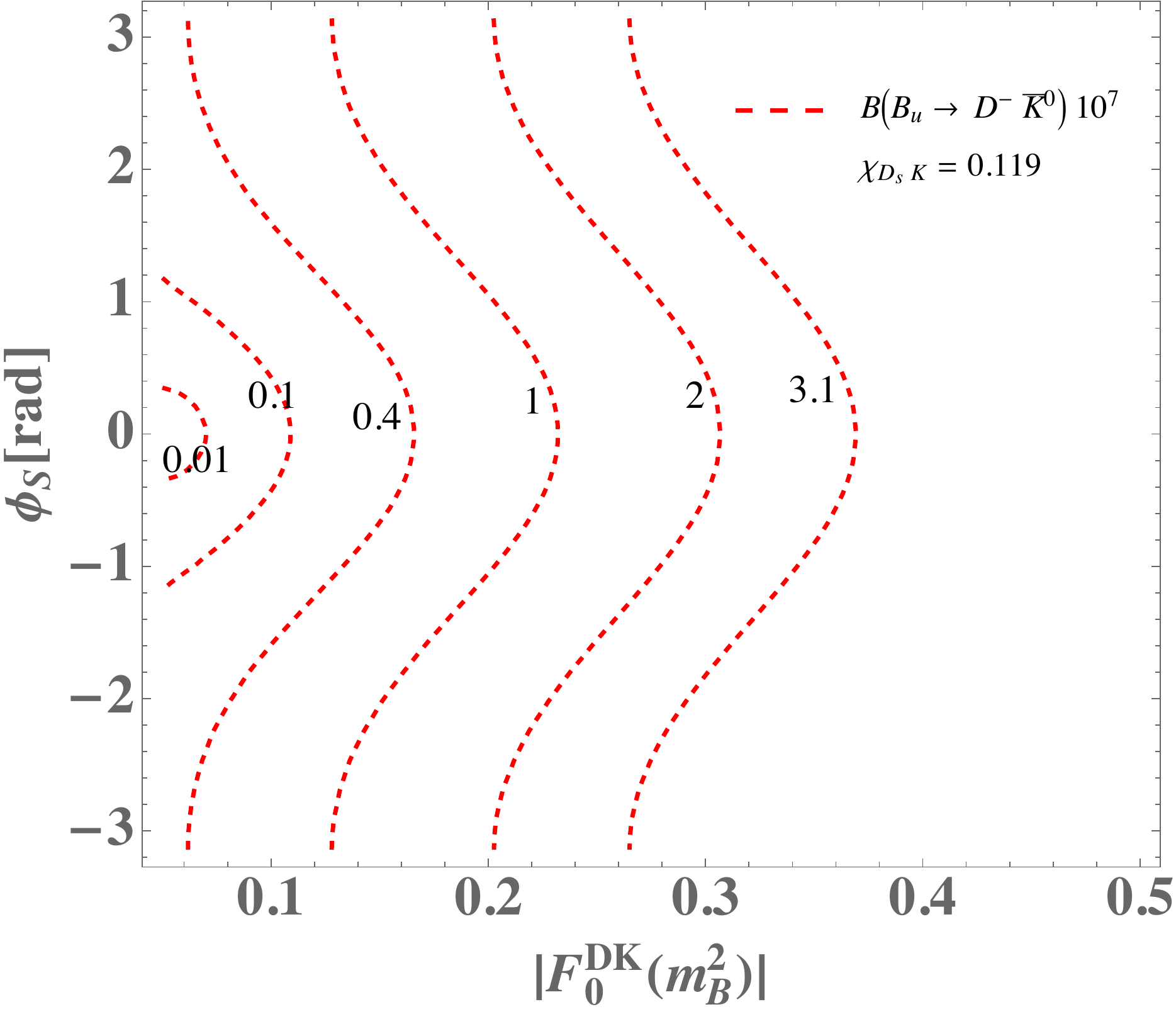}
 \caption{ Contours for ${\cal B}(B^-_u \to D^- \bar K^0)$ as a function of $|F^{DK}_0|$ and $\phi_S$. }
\label{fig:Cont_BDK}
\end{figure}

 The time-like form factor $F^{DK}_0(m^2_B)$ has not yet determined. Although the $B_u \to D^- \bar K^0$ decay is not observed,  we could use ${\cal B}(B^-_u \to D^- \bar K^{0})<3.1 \times 10^{-7}$, which was obtained earlier, to bound the magnitude of $F^{DK}_0(m^2_B)$. Using Eq.~(\ref{eq:ADKT}), the BR of $D^- K^0$  mode can be formulated as:
 \begin{align}
{\cal B}_{D^- K^0} &=  \tau_{B_u}\frac{G^2_F |V^*_{cs} V_{ub}|^2}{32 \pi m_B} \sqrt{\lambda_{DK}\left(\frac{m^2_D}{m^2_B}, \frac{m^2_K}{m^2_B} \right)} f^2_B \nonumber \\
& \times \left| a_1  (m^2_D -m^2_K) e^{i \phi_S} |F^{DK}_{0}(m^2_B)| - \frac{f_D C_1}{f_{D_s} N_c}  q^2  |\chi_{D_s K}(m^2_B)|\right|^2\,, \\
\lambda_{DK}(x,y) &=1 + x^2 + y^2 -2 x - 2 y - 2 x y \,, \nonumber 
\end{align}
where $\phi_S$ is the relative strong phase between $F^{DK}_0\equiv F^{DK}_0(m^2_B)$ and $\chi_{DK}\equiv \chi_{DK}(m^2_B)$. Using $|\chi_{D_s K}(m^2_B)|=0.119$, $m_{D}=1.870$ GeV, $m_{D_s}=1.968$ GeV, and the values in Table~\ref{tab:WCs},  the contours for ${\cal B}(B^-_u \to D^- \bar K^0)$ (in units of $10^{-7}$) as a function of $|F^{DK}_0(m^2_B)|$ and $\phi_S$ are shown in Fig.~\ref{fig:Cont_BDK}. From the plot, it can be clearly seen that the BR of $B^-_u\to D^- \bar K^0$ strongly depends on  $F^{DK}_0$ and the relative sign of $F^{DK}_0$ and $\chi_{DK}$.  The values of ${\cal B}(B^-_u \to D^- \bar K^0)$ with some benchmarks of $F^{DK}_0$ and $\phi_S$ are shown in Table~\ref{tab:BMF0phi}.  According to the analysis, we see that when $F^{DK}_0\sim 0.055$, the factorizable part of $A^{DK}_T$ becomes dominant.  

 %%%%

\begin{table}[htp]
\caption{ Branching ratio for $B^-_u \to D^- \bar K^0$ with some benchmarks of $|F^{DK}_0|$ and $\phi_S$. }
\begin{tabular}{c|ccccccc} \hline\hline
( $|F^{DK}_0|,~\phi_S$) & ~$(0,~0)$~& ~(0.20,~0)~& ~(0.20,~$2\pi/3$)~& ~(0.20,~$\pi$)~& ~(0.24,~0)~ &~(0.24,~ $2\pi/3$)~& ~(0.24,~$\pi$)~ \\ 
BR$(10^{-7})$ &  0.084  &  0.674 & 1.64 & 1.96 & 1.09  & 2.24 &  2.63 
 \\ \hline \hline

\end{tabular}
\label{tab:BMF0phi}
\end{table}%

 It is of interest to examine the rationality of our approach by comparing the BRs of $\bar B_d \to K^+ K^-$ and $\bar B_d \to D_s K^-$, in which both decays are from the $E_T$ topology.  Based on the decay-amplitude parametrizations given in Eqs.~(\ref{eq:MBKK}) and (\ref{eq:DKMM}), the ratio of branching fractions of $\bar B_d\to D_s K^-$ and $\bar B_d \to K^+ K^-$ can be obtained and estimated as:
\begin{align}
\frac{{\cal B}_{D_s K^-}}{{\cal B}_{K^+ K^-}} & \sim  \frac{|V^*_{ud} V_{cb}|^2}{|V^*_{ud} V_{ub}|^2} \frac{f^2_{Ds}}{f^2_K}=285\,,
\end{align}
 where we have included the  decay constants of $D_s$ and the $K$ mesons to show the effects from  different mesons. This numerical result fits well with the current data:
\begin{align}
 \frac{{\cal B}^{\rm exp}_{D_s K^-}}{{\cal B}^{\rm exp}_{K^+ K^-}} & = (2.83 \pm 0.63)\times 10^{2}\,.
\end{align}

\section{  $B^-_c\to (K^- K^0, J/\psi \pi^-)$, $B^-_c \to (\bar D^0 K^-, D^- \bar K^0)$, and $B^-_c \to \bar D^0 \pi^-$ decays}

  Analyzing the $B\to KK$, $B^-_u \to D^- \bar K^0$, and $\bar B_d \to D_s K^-$ decays, we can determine the nonfactorizable effect of the annihilation flavor diagram for the $B_u \to D^- \bar K^0$ decay. In addition, we can give a bound on the factorizable part of the same annihilation process. Based on the isospin symmetry,  we apply the obtained results  to the $B^-_c \to \bar D^0 K^-$ decay in this section. With a similar approach, we  estimate the BRs and CPAs for $B^-_c \to (D^-  \bar K^0,  \bar D^0 \pi^-)$. 

 Before investigating the $B^-_c \to \bar D^0 K^-$ decay,  we first apply our approach to predict ${\cal B}(B^-_c \to K^- K^0)$. According to the LHCb result of $R_{DK/J/\psi \pi}\approx 0.13 \pm 0.04$, if ${\cal B}(B^-_c \to J/\psi \pi^-)$ is known, we then have a clearer understanding of the BR for $B^-_c \to \bar D^0 K^-$. Therefore,  based on the $B_c \to J/\Psi$ form factor from lattice QCD~\cite{Colquhoun:2016osw}, we also estimate the BR for $B^-_c \to J/\psi \pi^-$.

\subsection{ $B^-_c \to K^- K^0$ and $B^-_c \to J/\psi \pi^-$}

It has been determined that the hadronic effect in  $\bar B_d \to K^- K^+$ is dominated by the nonfactorization contribution, and its effect can be directly related to the tree-annihilation  of  $B^-_u \to K^-  K^0$. The $B^-_c \to K^- K^0$ decay is dictated by the tree-annihilation diagram $A_T$, which is similar to that in $B^-_u \to K^-  K^0$; thus, we can estimate the BR for $B^-_c \to K^- K^0$ through the ${\cal B}(\bar B_d \to K^- K^+)^{\rm exp}$.  Using the parametrization defined in Eq.~(\ref{eq:TLNF}), the decay amplitudes for $\bar B_d \to K^- K^+$ and $B^-_c \to K^- K^0$ can be expressed as:
 \begin{align}
 M(\bar B_d \to K^- K^+) & \approx i \frac{G_F}{\sqrt{2}} V^*_{ud}V_{ub} \frac{C_2}{N_c} f_B q^2_1 \chi_{K^- K^+}(m^2_B)\,, \nonumber \\
 M(B^-_c \to K^- K^0) & \approx i \frac{G_F}{\sqrt{2}} V^*_{ud}V_{cb} \frac{C_1}{N_c} f_{B_c} q^2_2 \chi_{K^- K^0}(m^2_{B_c})
 \end{align}
with $q^2_{1(2)}\approx 4 m^2_K - m^2_{B(B_c)}$. If we take the asymptotic form factor behavior as $\chi_{KK}(Q^2) \propto 1/Q^2$, the ratio of the branching fraction of $B^-_c \to K^- K^0$ to $\bar B_d \to K^- K^0$ can be obtained as
 \begin{equation}
 \frac{{\cal B}(B^-_c \to K^- K^0)}{{\cal B}(\bar B_d \to K^- K^+)} \approx \frac{\tau_{B_c} m_{B}}{\tau_{B_d} m_{B_c}} \frac{|C_1 V_{cb}|^2}{|C_2 V_{ub}|^2} \frac{f^2_{B_c}}{f^2_B}\approx 8.96\,,
 \end{equation}
 where $\tau_{B_c}=0.507$ ps and  $f_{B_c}=0.434$ GeV are used~\cite{Colquhoun:2015oha}. With ${\cal B}(\bar B_d \to K^- K^+) =7.8 \times 10^{-8}$, the BR for $B^-_c \to K^- K^0$ is ${\cal B}(B^-_c \to K^- K^0) \approx 6.99 \times 10^{-7}$, where the result is a factor of 2.9  larger than the estimation in the perturbative QCD (PQCD) approach~\cite{Liu:2009qa}. 
 
 The $B^-_c \to J\psi \pi^-$ decay is a color-allowed tree process. Since the nonfactorization effect is related  to $C_1/N_c$, it is expected that the factorization effect will dominate.   Although $J/\psi$ is a vector-boson, only longitudinal polarization has a  contribution in $B^-_c \to J/\psi \pi^-$; thus, the decay amplitude with the factorizable part can be written as;
 \begin{equation}
 M(B^-_c \to J/\psi \pi^-) \approx  -\frac{G_F}{\sqrt{2}} V^*_{ud} V_{cb} a_1 \left[2 m_{J/\psi} f_\pi A^{B_c J/\psi}_0(m^2_\pi) \right] \varepsilon^*_{J/\psi} \cdot p_\pi\,,
 \end{equation}
where $A^{B_c J/\psi}_0$ is one of $B_c \to J/\psi$ transition form factors. Accordingly, with the approximation of $m^2_\pi/m^2_{B_c} \approx 0$, the BR for $B_c \to J/\psi \pi^-$ can be formulated as:
 \begin{equation}
 {\cal B}(B^-_c \to J/\psi \pi^-) \approx \tau_{B_c} \frac{G^2_F m^3_{B_c} | V^*_{ud} V_{cb}|^2}{32 \pi} \left(a_1 f_\pi A^{B_c J/\psi}_0(m^2_\pi) \right)^2\,. \label{eq:BcJpsipi}
 \end{equation}
The unknown in Eq.~(\ref{eq:BcJpsipi}) is $A^{B_c J/\psi}_0(m^2_\pi)\approx A^{B_c J/\psi}_0(0)$. The  form factor values calculated using various QCD approaches vary considerably in the literature~\cite{Du:1988ws,Colangelo:1992cx,Kiselev:1993ea,Nobes:2000pm,Ivanov:2000aj,Kiselev:2000pp,Ebert:2003cn,Ivanov:2005fd,Hernandez:2006gt,Huang:2007kb,Sun:2008ew,Dhir:2008hh,Wang:2008xt,Qiao:2011yz,Dubnicka:2017job}. Recently, the HPQCD collaboration has made  progress in the calculations of the form factors for  $B_c\to J/\psi$~\cite{Colquhoun:2016osw},  
and the obtained (preliminary) results are given as:
 \begin{align}
% & f_+ = [ a\pm b, a\pm b]\,, \  f_{-}=[a\pm b,\, a \pm b] \nonumber \\
 %
 A^{B_cJ/\psi}_1(0) \approx 0.49\,, \  V^{B_c J/\psi}(0)\approx 0.77\,, \label{eq:HPQCD}
 \end{align}
where the uncertainties of the form factors could be around $10\%$ or less. Taking the HPQCD results as  theoretical guidance, the results, which are calculated by QCD models and  are all within $10\%$ of  the values in Eq.~(\ref{eq:HPQCD})~\cite{Nobes:2000pm,Wang:2008xt}, indicate $A^{B_cJ/\psi}_0(0) \approx A^{B_cJ/\psi}_1(0)$. Thus, with the indication and $10\%$ uncertainty of $A^{B_cJ/\psi}_0(0)$, the BR of  $B_c\to J/\psi \pi^-$ can be obtained as:
 \begin{equation}
 {\cal B}(B^-_c\to J/\psi \pi^-) \approx  (7.7\pm 1.1)\times 10^{-4}\,.
 \end{equation}
Using above result and $R_{DK/J/\psi \pi}\approx 0.13 \pm 0.04$,  the BR of $B^-_c \to  \bar D^0 K^-$ can be  estimated as:
 \begin{equation}
 {\cal B}(B^-_c \to \bar D^0 K^-) \approx (10.01 \pm 3.40) \times 10^{-5}\,. \label{eq:VBRBcDK}
 \end{equation}

\subsection{Branching ratios and CP asymmetries for $B^-_c \to (\bar D^0  K^-, D^- \bar K^0)$}

\begin{figure}[phtb]
\includegraphics[scale=0.8]{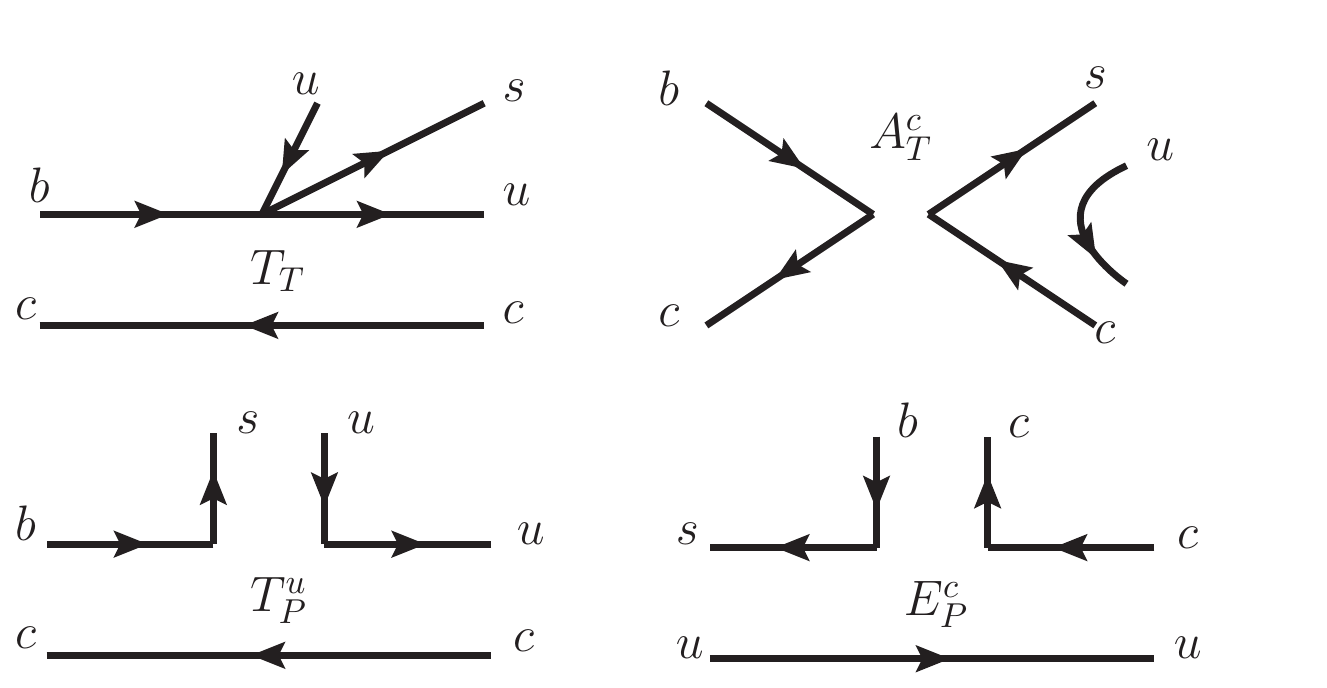}
 \caption{ Flavor diagrams for the $B^-_c \to \bar D^0 K^-$  decay. }
\label{fig:Topo_BcDK}
\end{figure}

Similar to the $B\to KK$ decays, the $B^-_c \to \bar D^0 K^-$ decay mainly arises from the gluonic penguin and $W$-mediated tree Feynman diagrams. The portion of the effective Hamiltonian  can be obtained from Eq.~(\ref{eq:HBKK}) by using $s$-quark instead of $d$-quark. In addition to the $(\bar u b)_{V-A} (\bar s u)_{V-A}$ operator,  the $(\bar c b)_{V-A} (\bar s c)_{V-A}$ operator is also involved in the $B^-_c \to \bar D^0 K^-$ decay, where the corresponding Hamiltonian can be obtained from that for $B^-_u \to D^- \bar K^0$, as shown in  Eq.~(\ref{eq:HDK}), by replacing $u$-quark with $c$-quark, i.e.,  $V_{ub} \to V_{cb}$ and $Q_{1,2} \to Q^{c}_{1,2}$. Accordingly, the topological  flavor diagrams for $B^-_c \to \bar D^0 K^-$ are shown in Fig.~\ref{fig:Topo_BcDK}. 

 From the flavor diagrams,  the decay amplitude for $B^-_c \to \bar D^0 K^-$ can be written as:
 \begin{equation}
 M(B^-_c \to \bar D^0 K^-) = \frac{G_F}{\sqrt{2}} \left[V^*_{us} V_{ub}  T_{T} +  V^*_{cs} V_{cb} A^c_T  - V^*_{ts} V_{tb} \left( T^u_P + E^c_P \right) \right]\,, \label{eq:BcDK}
 \end{equation}
where $T^u_P$ and $E^c_P$ are similar to $T^{q'}_P$ and $E^{q'}_P$ in $\bar B_d \to \bar K^0 K^0$ and $B^-_u \to K^- K^0$, respectively, $A^c_T$ is similar to $A^{DK}_T$ in $B^-_u \to D^- \bar K^0$, and $T_T$ denotes the contribution from the tree transition topology. Since $T_T$ is dominated by the color-allowed effect and the associated WC is $a_1\sim 1$, it is expected that $T_T$ will be predominantly dictated by the factorizable part. A similar situation is also suitable for $T^u_P$. In order to describe $T_T$ and $T^u_P$, we need the $B_c \to \bar D^0$ transition form factors, which are defined as:
 \begin{equation}
 \langle \bar D^0 (P'_2)| \bar u \gamma_\mu b |  B^-_c (P'_1)\rangle = f^{B_c D}_1(q^2) \left( P_\mu -\frac{P\cdot q}{q^2} q_\mu  \right) + f^{B_c D}_0 (q^2) \frac{P\cdot q}{q^2} q_\mu\,,
 \end{equation}
where $P=P'_1 + P'_2$, $q=P'_1-P'_2$, $P\cdot q=m^2_{B_c} -m^2_{D}$, and $f^{B_cD}_{1,0}$ are the form factors. As a result, we obtain $\langle \bar D^0 | \bar u \gamma_\mu b | B^-_c\rangle q^\mu= (m^2_{B_c} -m^2_D) f^{B_c D}_0(m^2_K)$ and $\langle \bar D^0 | \bar u  b | B^-_c\rangle = (m^2_{B_c} -m^2_D)/(m_b -m_u) f^{B_c D}_0(m^2_K)$.  Thus, with the factorization assumption, $T_T$ and $T^u_P$ can be expressed as:
 \begin{align}
 T_T &=\langle \bar D^0 K^- | C_1 Q'^c_1 + C_2 Q'^c_2 | B^-_c \rangle \approx -i a_1f_K (m^2_{B_c} -m^2_D) f^{B_c D}_0(m^2_K)\,, \\
  T^u_P & = \langle \bar D^0 K^- | \sum^6_{i=3} C_i O_i | B^-_c \rangle \approx -i a_{46} f_K (m^2_{B_c} -m^2_{D} ) f^{B_cD}_0(m^2_K)\,, \\
  a_{46} & = a_4 + 2a_6 \frac{m^2_K}{(m_s+m_u)(m_b-m_u)}    \,. \nonumber 
 \end{align}
 According to  Eqs.~(\ref{eq:EuP}) and (\ref{eq:ADKT}),  $A^c_T$ and $E^c_P$ can be parametrized as:
 \begin{align}
 A^c_T & \approx - i  f_{B_c} \left[  a_1 (m^2_D -m^2_K) e^{i \phi_S}  |F^{DK}_0(m^2_{B_c})| -  \frac{C_1}{ N_c}   q^2_{B_c}  |\chi_{D K}(m^2_{B_c})|\right]\,, \nonumber \\
 %
 %T^u_P & \approx  -i   \left(a_4 + 2a_6 \frac{m^2_K}{(m_s+m_u)(m_b-m_u)}  \right) f_K (m^2_{B_c} -m^2_{D} ) f^{B_cD}
 %_0(m^2_K) \,, \nonumber \\
 %
 E^c_P & \approx i 2 a_6 \frac{f_{B_c} m^2_{B_c} }{m_b + m_c}\frac{m^2_D -m^2_K}{m_c -m_u} e^{i\phi_S}|F^{DK}_0 (m^2_{B_c})|
 \end{align}
where $q^2_{B_c}=  2(m^2_{D} + m^2_K) -m^2_{B_c}$, and $\chi_{DK}(m^2_{B_c}) \approx f_D/f_{D_s} \chi_{D_s K}(m^2_{B_c})$. Taking the asymptotic behaviors of $F^{DK}_0(Q^2)$ and $\chi_{D K}(Q^2)$ as $\propto1/Q^2$, we get $F^{DK}_0(m^2_{B_c})\approx m^2_B/m^2_{B_c} F^{DK}_0(m^2_B)$ and $\chi_{DK}(m^2_{B_c}) \approx f_D m^2_{B} /(f_{D_s} m^2_{B_c}) \chi_{D_s K}(m^2_{B})$.

\begin{figure}[phtb]
\includegraphics[scale=0.35]{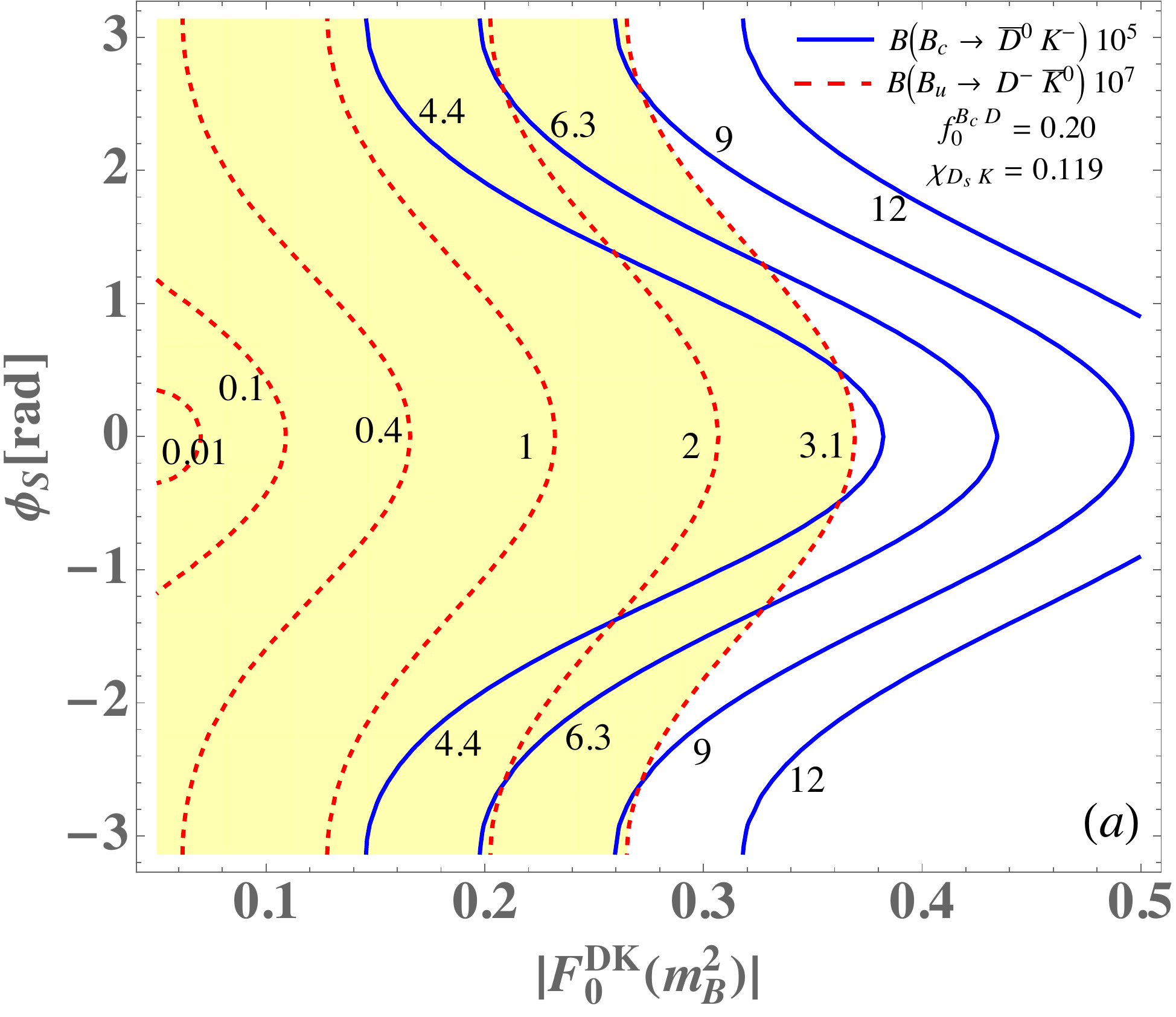}
\includegraphics[scale=0.35]{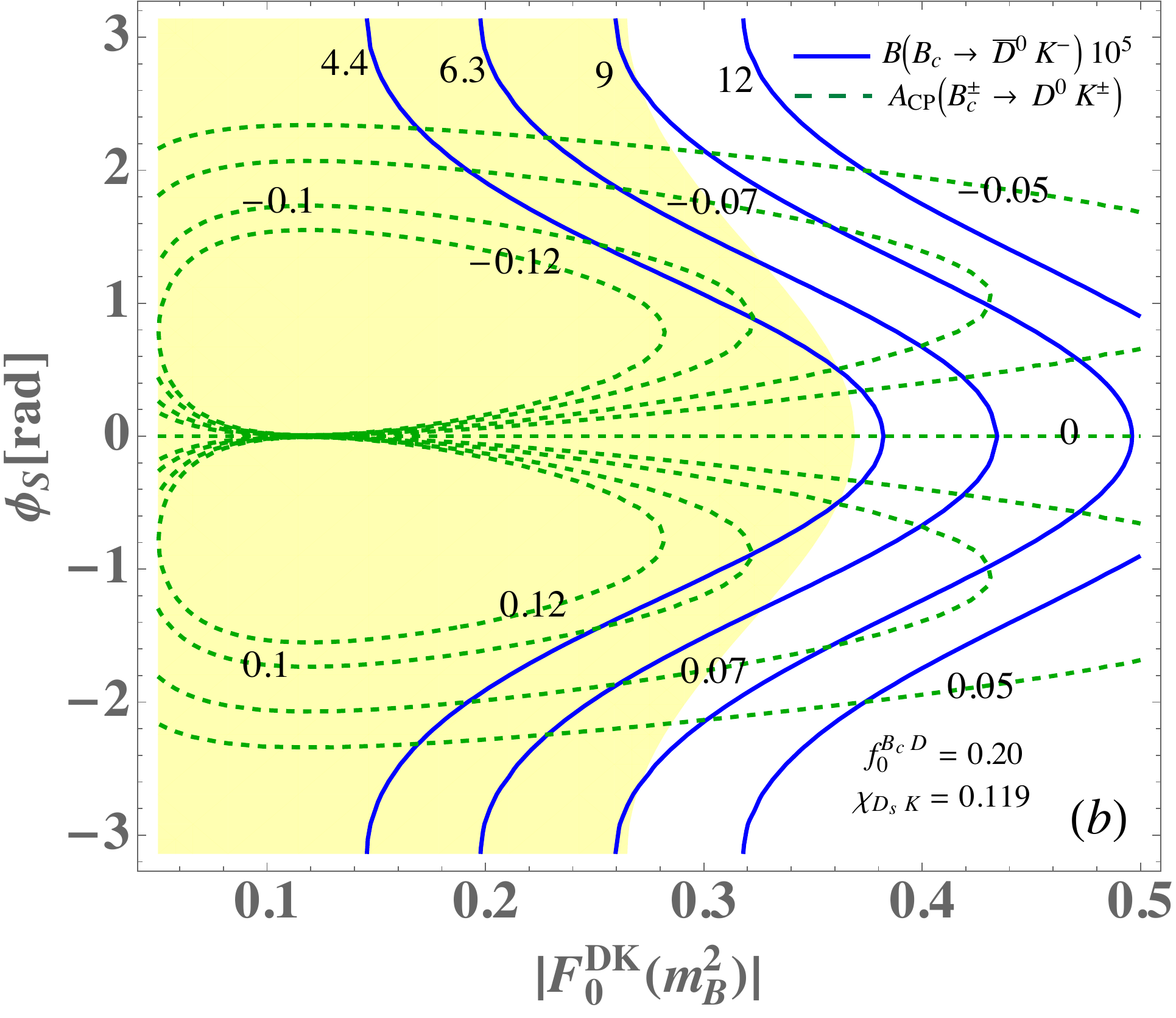}
 \caption{(a) Contours for (a) ${\cal B}^{\rm avg}_{D^0 K^\pm}$ (solid, in units of $10^{-5}$) and (b)  $A_{CP}( B^\pm_c \to  D^0 K^\pm)$ (dashed) as a function of $\phi_S$ and $|F^{DK}_0(m^2_B)|$, where we have fixed $f^{B_cD}_0=0.20$ and $\chi_{D_s K}=0.119$. For comparison, we also show ${\cal B}(B^-_u \to D^- \bar K^0)$ (dashed, in units of $10^{-7}$) in (a).  The shaded area denotes the region of ${\cal B}(B^-_u \to D^- \bar K^0)<3.1 \times 10^{-7}$.}
\label{fig:BRACP}
\end{figure}

Except the strong phase $\phi_S$ and form factors $F^{DK}_0(m^2_B)$ and $f^{B_c D}_0(m^2_K)$, basically, we have most of the information necessary  to calculate the BR and CPA for $B^-_c \to \bar D^0 K^-$, which are defined as:
 \begin{align}
 {\cal B}(B^-_c \to \bar D^0 K^-) &= \frac{\tau_{B_c}}{16 \pi m_{B_c}} \sqrt{\lambda_{DK}\left( \frac{m^2_D}{m^2_{B_c}},\frac{m^2_K}{m^2_{B_c}}\right)} \left| M(B^-_c \to \bar D^0 K^-)\right|^2\,,  \label{eq:BRBcDK} \\
 A_{CP}(B^\pm_c \to D^0 K^\pm)&=\frac{{\cal B}(B^-_c \to \bar D^0 K^-) - {\cal B}(B^+_c \to  D^0 K^+) }{{\cal B}(B^-_c \to \bar D^0 K^-) +{\cal B}(B^+_c \to  D^0 K^+) }\,. \label{eq:ACPBcDK}
 \end{align}
The BR of CP-average can be obtained via $ {\cal B}^{\rm avg}_{ D^0 K^\pm} =  [{\cal B}(B^-_c \to \bar D^0 K^-)  +  {\cal B}(B^+_c \to D^0 K^\pm)]/2$. The contour plots for $ {\cal B}^{\rm avg}_{D^0 K^\pm}$ (solid, in units of $10^{-5}$) and $A_{CP}(B^\pm_c \to  D^0 K^\pm)$ (dashed) as a function of $\phi_S$ and $|F^{DK}_0(m^2_B)|$ are respectively shown in Fig.~\ref{fig:BRACP}(a) and (b), where $f^{DK}_0(m^2_K)=0.20$ and $\chi_{D_s K}(m^2_B)=0.119$ are used; for comparison, we also show ${\cal B}(B^-_u \to D^- \bar K^0)$ (dashed, in units of $10^{-7}$) in Fig.~\ref{fig:BRACP}(a),  where the shaded area denotes the range of ${\cal B}(B^-_u \to D^- \bar K^0)<3.1 \times 10^{-7}$. The contour lines  marked as $(4.4, 6.3) \times 10^{-5}$ denote the values  taken from the downward $1.5\sigma$ and $1\sigma$  of Eq.~(\ref{eq:VBRBcDK}).  Based on our analysis, it can be seen that  with $f^{B_c D}_0(m^2_K)=0.20$, the ${\cal B}(B^-_u \to D^- \bar K^0)$ has to be larger (less) than $1(3.1)\times 10^{-7}$ when ${\cal B}(B^-_c \to  \bar D^0 K^-) > 4.4 \times 10^{-5}$, and due to the upper bound of $B^-_u \to D^- \bar K^0$, the BR of $B^-_c \to  \bar D^0 K^-$ should be less than approximately $9\times 10^{-5}$. Since $T^u_P$ is proportional to $f^{B_cD}_0$, it is expected that with a larger value of $f^{B_cD}_0$, the curves for ${\cal B}^{\rm avg}_{D^0 K^\pm}$ in Fig.~\ref{fig:BRACP}(a)  will shift to the left; that is, a larger ${\cal B}^{\rm avg}_{D^0 K^\pm}$ is allowed. Since the calculation results of $f^{B_c D}_0$ are quite diverse and spread from $0.075$ to $0.69$~\cite{Du:1988ws,Colangelo:1992cx,Nobes:2000pm,Ivanov:2000aj,Kiselev:2000pp,Ebert:2003cn,Huang:2007kb,Dhir:2008hh,Wang:2008xt,Dubnicka:2017job}, we need the input from the lattice calculations to determine the more accurate form factor.  If we take the HPQCD calculations on $B^-_c \to (\eta_c, J/\psi)$ as a guide, the result from the light-front QCD model, where the predicted form factors of $B^-_c \to (\eta_c, J/\psi)$ fall within $10\%$ of the HPQCD results, prefers  $f^{B_cD}_0 \sim 0.18$ with $f_{B_c}=0.440$ GeV~\cite{Wang:2008xt}.  The  value of $f^{B_c D}_0 =0.2$ used in our analysis fits the preference and is comparable with the result in~\cite{Zhang:2009ur} using the PQCD approach.  Although we cannot precisely predict the CPA, from Fig.~\ref{fig:BRACP},  its magnitude should be $|A_{CP}(B^\pm_c \to  D^0 K^\pm| \lesssim 10\%$. The CPA can be up to $30\%$ if $f^{B_cD}_0 = 0.60$ is used. 

\begin{figure}[phtb]
\includegraphics[scale=0.35]{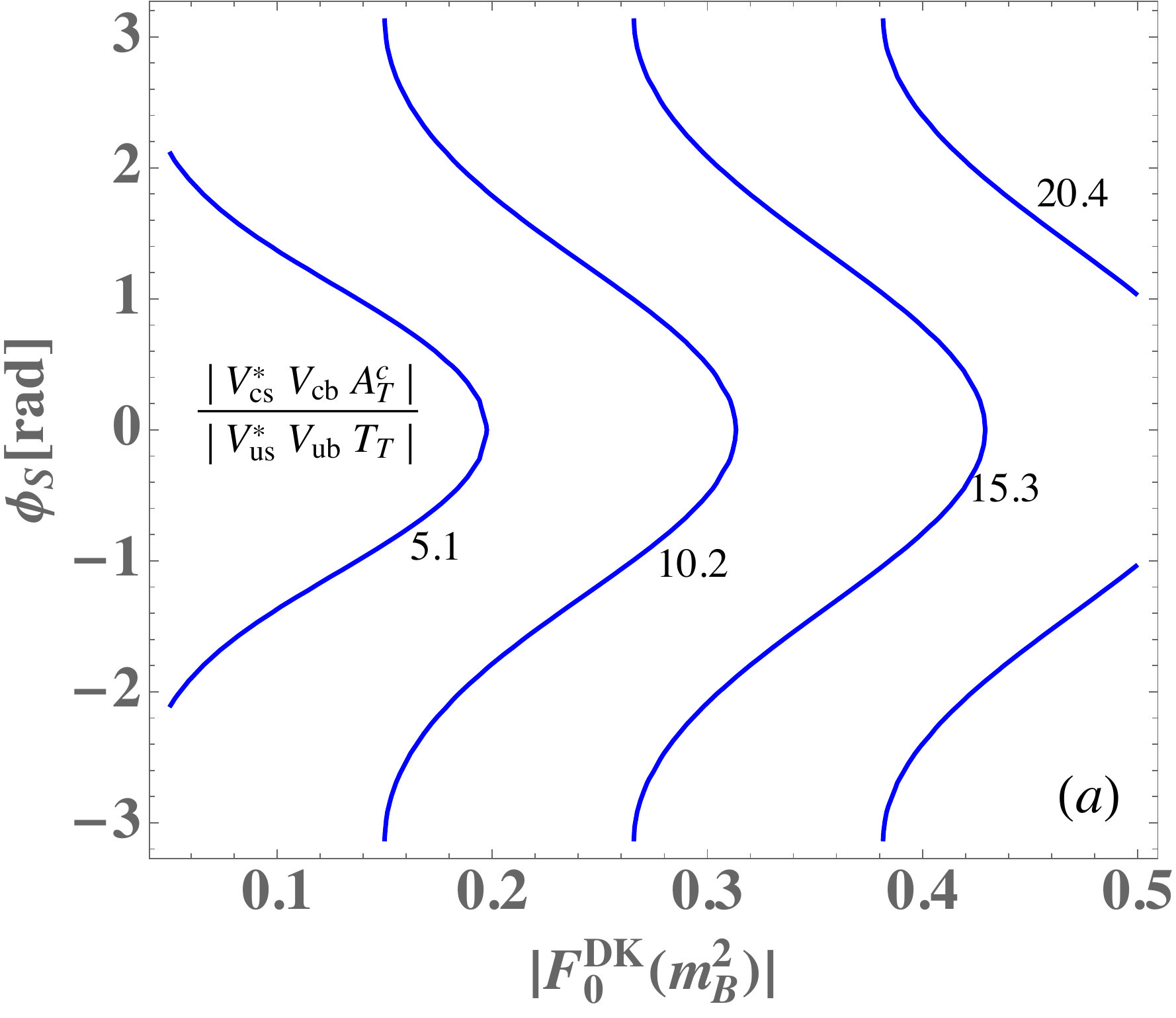}
\includegraphics[scale=0.35]{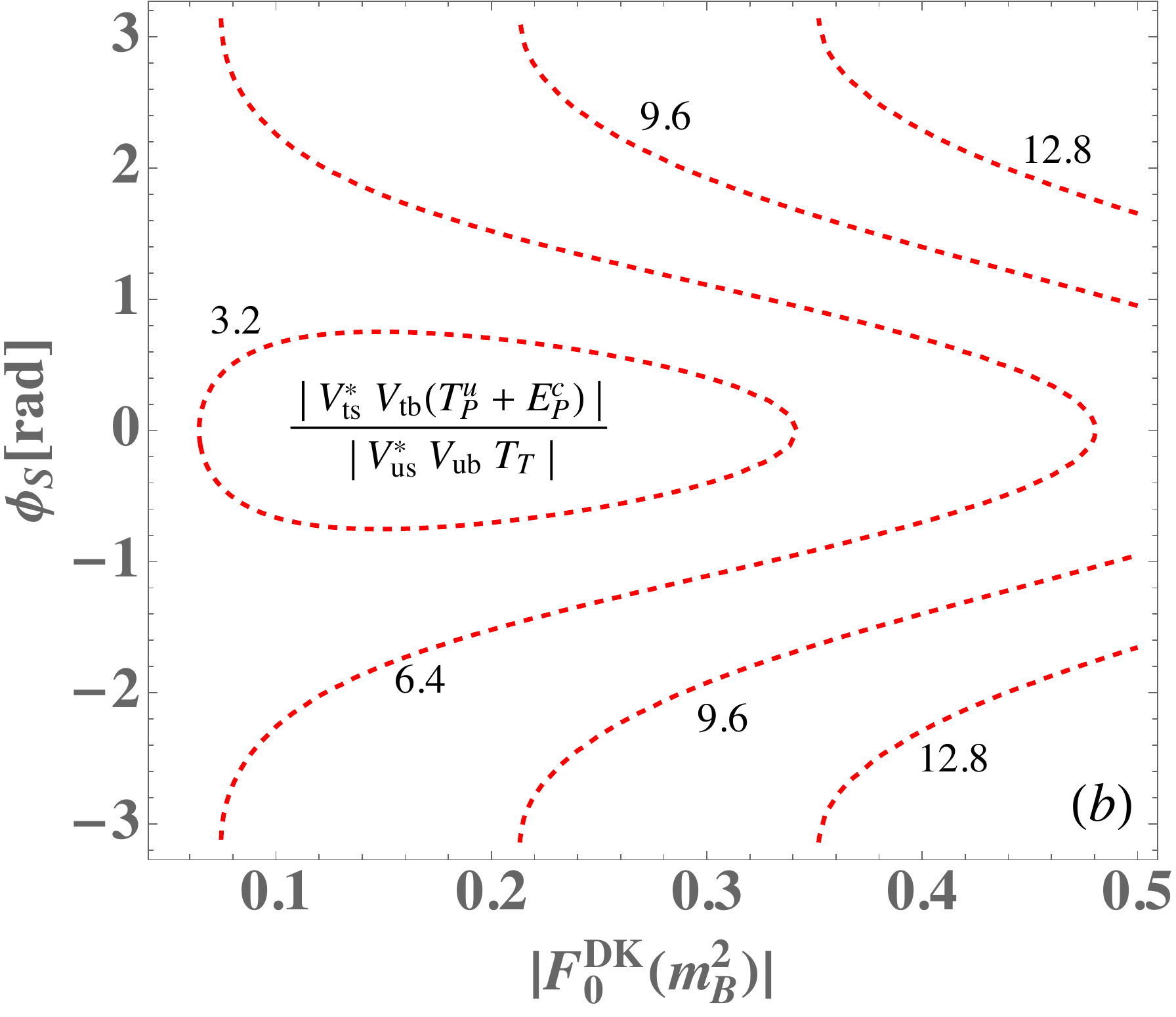}
\includegraphics[scale=0.35]{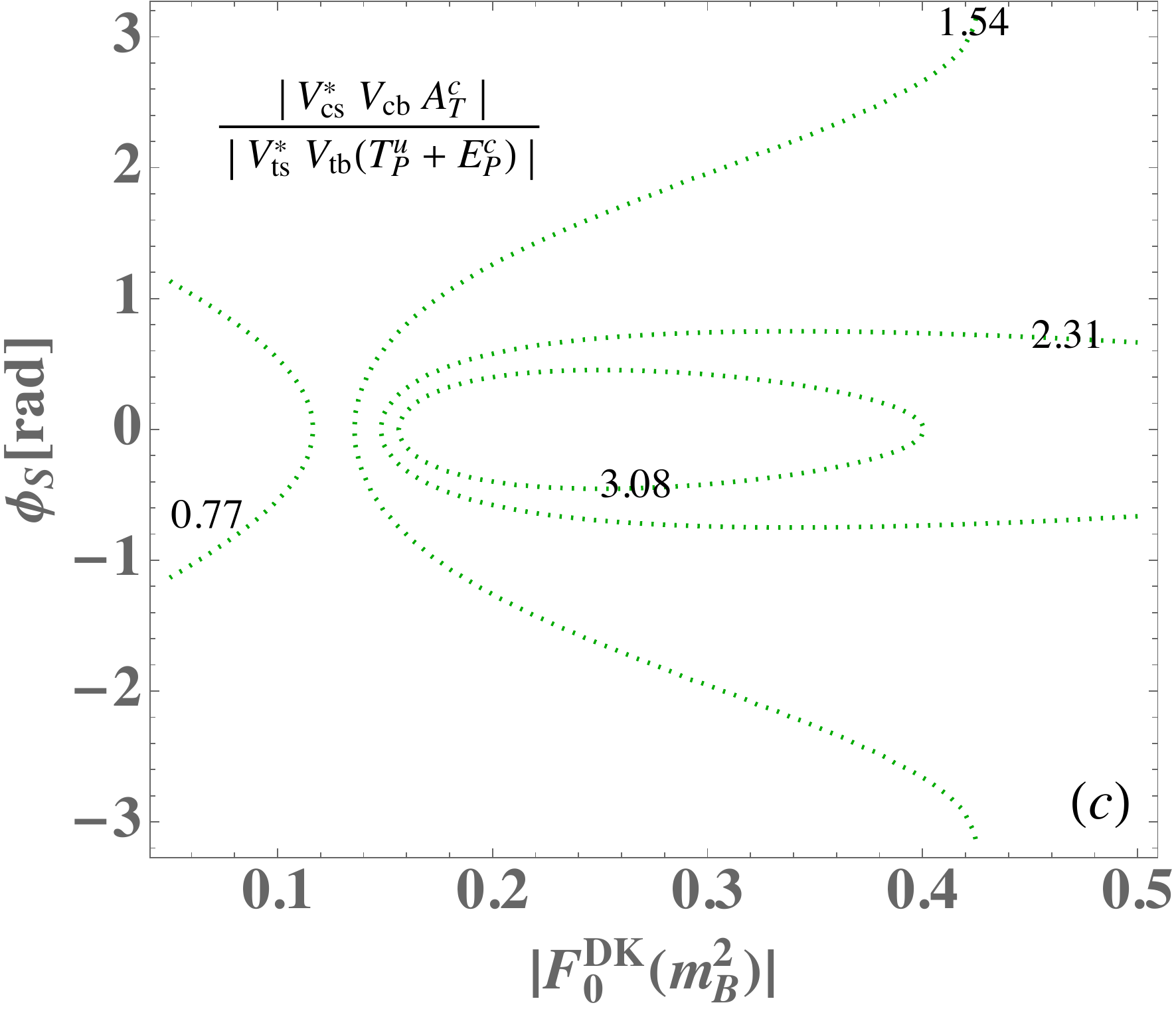}
\includegraphics[scale=0.35]{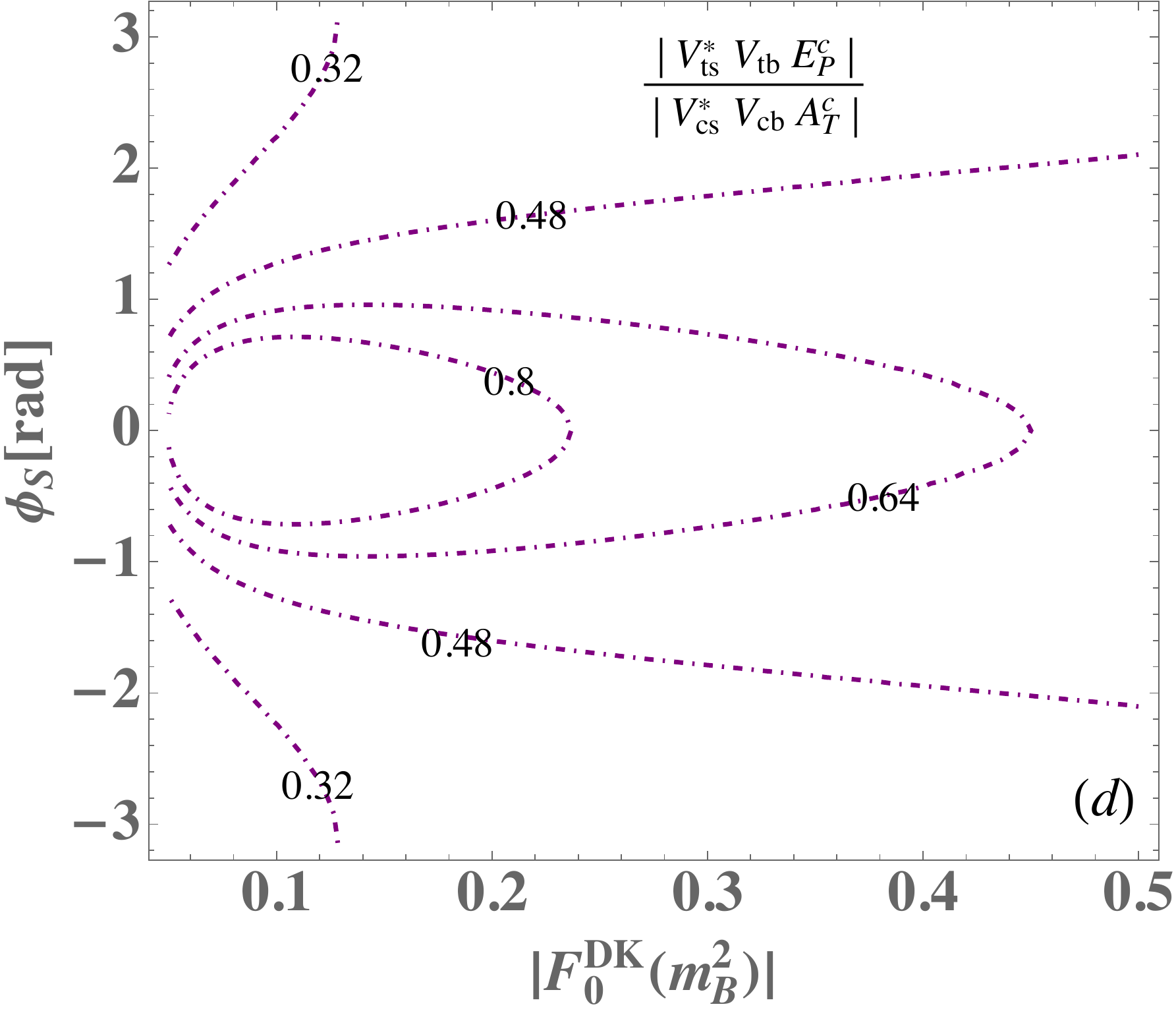}
 \caption{Contours for (a) $|V^*_{cs} V_{cb} A^c_T|/|V^*_{us} V_{ub} T_T|$, (b) $|V^*_{ts} V_{tb} (T^u_P+E^c_P)|/|V^*_{us} V_{ub} T_T|$, (c) $|V^*_{cs} V_{cb} A^c_T|/|V^*_{ts} V_{tb} (T^u_P+E^c_P)|$, and (d) $|V^*_{ts} V_{tb} E^c_P|/|V^*_{cs} V_{cb} A^c_T|$ as a function of $\phi_S$ and $|F^{DK}_0(m^2_B)|$, where we take $f^{B_cD}_0=0.20$. }
\label{fig:Ra-d}
\end{figure}

It was mentioned earlier that although \cite{Du:1998te} and \cite{Zhang:2009ur}  using the different QCD approaches  can obtain ${\cal B}(B^-_c \to  \bar D K^-) \sim 5\times 10^{-5}$, their origins related to enhancing  the BR are different. In order to show the role of each component in the decay amplitude shown in Eq.~(\ref{eq:BcDK}), we show the ratios of $|V^*_{cs} V_{cb} A^c_T|/|V^*_{us} V_{ub} T_T|$, $|V^*_{ts} V_{tb} (T^u_P+E^c_P)|/|V^*_{us} V_{ub} T_T|$, $|V^*_{cs} V_{cb} A^c_T|/|V^*_{ts} V_{tb} (T^u_P+E^c_P)|$, and $|V^*_{ts} V_{tb} E^c_P|/|V^*_{cs} V_{cb} A^c_T|$ in Fig.~\ref{fig:Ra-d}(a)-(d), where $f^{B_cD}_0=0.20$ is used. From plots (a) and (b), it can be clearly seen that the tree-annihilation and penguin effects offer the dominant contributions. We can further see  from plot (c) that  when ${\cal B}(B^-_c \to \bar D^0 K^-) > 4.4 \times 10^{-5}$, the contribution from the tree-annihilation  is larger than that from the penguin topologies.  According to plot (d), it is known that the penguin-annihilation topology $E^c_P$ is smaller than the tree-annihilation topology $A^c_T$. Hence, our results are consistent with~\cite{Zhang:2009ur}. 

Now, we can  apply all calculations to the $B^-_c \to D^- \bar K^0$ decay, where the effective Hamiltonian is the same as that for the $B^-_c \to \bar D^0 K^-$ decay. It can be easily found that with the exception of the $T_T$ topology diagram, which does not appear in $B^-_c \to D^- \bar K^0$,  the decay amplitude of $B_c \to D^- \bar K^0$ can be obtained from that in Eq.~(\ref{eq:BcDK}) by replacing $u$-quark with $d$-quark. With the isospin symmetry, we can write the decay amplitude as:
 \begin{equation}
 M(B^-_c \to D^- \bar K^0) = M(B^-_c \to \bar D^0 K^-) - \frac{G_F}{\sqrt{2}} V^*_{us} V_{ub} T_T\,.
 \end{equation}
The BR for $B^-_c \to D^- \bar K^0$ can be calculated using Eq.~(\ref{eq:BRBcDK}).  As shown before, the $B^-_c \to \bar D^0 K^-$  decay is dominated by the tree-annihilation and penguin topologies; thus, we expect ${\cal B}(B^-_c \to D^- \bar K^0) \approx {\cal B}(B^-_c \to D^- \bar K^0)$. Since  \cite{Du:1998te} took a larger $f^{B_c D}_0\sim 0.6$ and got ${\cal B}(B^-_c \to D^- \bar K^0)/{\cal B}(B^-_c \to D^- \bar K^0)\approx 0.75$, we can use the different predictions of ${\cal B}(B^-_c \to D^- \bar K^0)$ to test the different approaches. The BR values for $B^-_c \to D^- \bar K^0$ with the same benchmarks shown in Table~\ref{tab:BMF0phi} are given in Table~\ref{tab:BRBcDnK0}. Due to the small  weak CP violating phase in $V_{ts}$,  the CPA of $B^-_c \to D^- \bar K^0$ is suppressed.  

\begin{table}[htp]
\caption{ Branching ratio for $B^-_c \to D^- \bar K^0$ with the benchmarks shown in Table~\ref{tab:BMF0phi}, where $f^{B_cD}_0=0.20$ is used.  }
\begin{tabular}{c|ccccccc} \hline\hline
( $|F^{DK}_0|,~\phi_S$) & ~$(0,~0)$~& ~(0.20,~0)~& ~(0.20,~$2\pi/3$)~& ~(0.20,~$\pi$)~& ~(0.24,~0)~ &~(0.24,~ $2\pi/3$)~& ~(0.24,~$\pi$)~ \\ 
BR$(10^{-5})$ &  0.96  & 0.37  & 5.03 & 6.58 &  0.86   & 6.45 &  8.3 
 \\ \hline \hline

\end{tabular}
\label{tab:BRBcDnK0}
\end{table}%

\subsection{  Predictions of the $B^-_c \to \bar D^0 \pi^-$ decay}

\begin{figure}[phtb]
\includegraphics[scale=0.35]{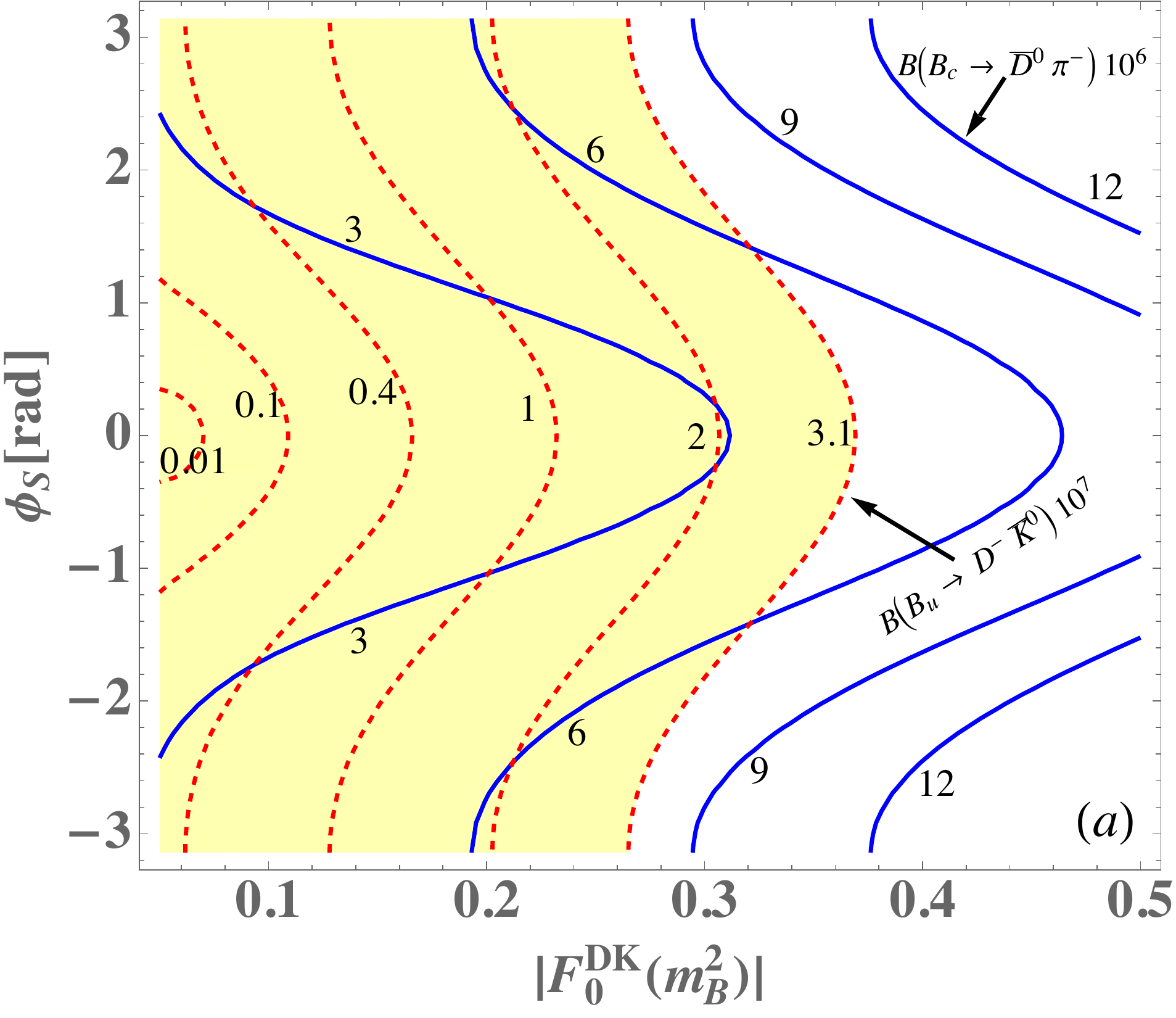}
\includegraphics[scale=0.35]{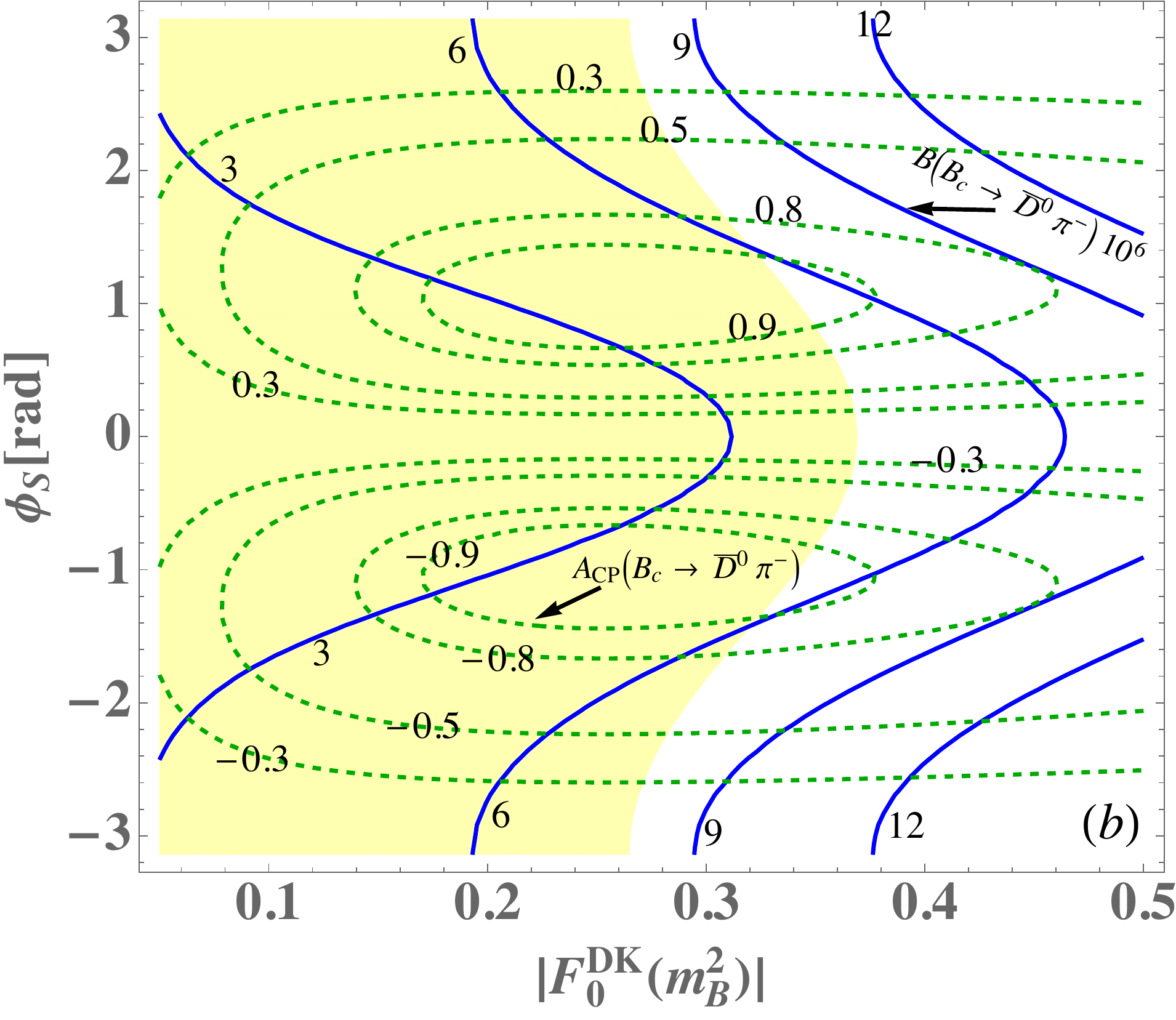}
 \caption{The legend is the same as Fig.~\ref{fig:BRACP} but for $B^-_c \to  \bar D^0 \pi^-$. }
\label{fig:BRACPpi}
\end{figure}

The $B^-_c \to \bar D^0  \pi^-$ decay is of interest because apart from the CKM matrix elements, it has very similar topological flavor diagrams as those in the $B^-_c \to \bar D^0 K^-$ decay. When the $s$-quark in Eq.~(\ref{eq:BcDK}) is replaced by the $d$-quark,   the decay amplitude for $B^-_c \to \bar D^0 \pi^-$ can be written as:
 \begin{equation}
 M(B^-_c \to \bar D^0 \pi^-) = \frac{G_F}{\sqrt{2}} \left[V^*_{ud} V_{ub}  T'_{T} +  V^*_{cd} V_{cb} A'^c_T  - V^*_{td} V_{tb} \left( T'^u_P + E'^c_P \right) \right]\,. \label{eq:BcDpi}
 \end{equation}
The hadronic effects $T'_T$, $A'^c_T$, $T'^u_P$, and $E'^c_P$ are given as:
  \begin{align}
  T'_T &=\langle \bar D^0 \pi^- | C_1 O_1 + C_2 O_2 | B^-_c \rangle \approx -i a_1f_\pi (m^2_{B_c} -m^2_D) f^{B_c D}_0(m^2_\pi)\,, \nonumber \\ 
  T'^u_P & = \langle \bar D^0 \pi^- | \sum^6_{i=3} C_i O_i | B^-_c \rangle \approx -i a'_{46} f_\pi (m^2_{B_c} -m^2_{D} ) f^{B_cD}_0(m^2_\pi)\,, \nonumber \\
A'^c_T & \approx - i  f_{B_c} \frac{f_\pi}{f_K}\left[  a_1  (m^2_D -m^2_K) e^{i \phi_S}  |F^{DK}_0(m^2_{B_c})| -  \frac{C_1}{ N_c}   q^2_{B_c}  |\chi_{D K}(m^2_{B_c})|\right]\,, \nonumber \\
 E'^c_P & \approx i 2 a_6 \frac{f_{B_c} m^2_{B_c} }{m_b + m_c}\frac{m^2_D -m^2_K}{m_c -m_u} \frac{f_\pi e^{i\phi_S}}{f_K}|F^{DK}_0 (m^2_{B_c})|\,,
  \end{align}
where we have included the SU(3) breaking effect $f_\pi /f_K$ for the form factors $F^{DK}_0$ and $\chi_{DK}$, and $a'_{46}  = a_4 + 2a_6 m^2_\pi/(m_d+m_u)/(m_b-m_u)$.  It can be seen that compared to $B^-_c \to \bar D^0 K^-$,  the tree-annihilation $A'^c_T$ has an extra Wolfenstein parameter suppression factor $\lambda\approx 0.22$  from $V^*_{cd}$; however, the $T'_T$ contribution is associated with $V^*_{ud}\sim 1$, which is $1/\lambda$ larger than $V^*_{us}$; that is, the tree-annihilation topology does not dominate anymore in this process.  Since the calculations for the BR and CPA of $B^-_c \to  \bar D^0 \pi^-$ are the same as those for $B^-_c \to  \bar D^0 K^-$,  based on Eqs.~(\ref{eq:BRBcDK}) and (\ref{eq:ACPBcDK}), the contours for the BR and CPA of $B^-_c \to  \bar D^0 \pi^-$ as a function of $\phi_S$ and $|F^{DK}_0(m^2_B)|$ are shown in Fig.~(\ref{fig:BRACPpi})(a) and (b), respectively. Due to the upper limit of $B^-_u \to D^- \bar K^0$, we obtain ${\cal B}^{\rm avg}_{ D^0 \pi^\pm} < 8\times 10^{-6}$. If we assume $4.4 < {\cal B}^{\rm avg}_{D^0 K^\pm}10^5 < 9.0$, the corresponding range for ${\cal B}^{\rm avg}_{D^0 \pi^\pm}$ is $4.9 < {\cal B}^{\rm avg}_{D^0 \pi^\pm}10^6 < 8$.  Since the CKM matrix elements of tree and penguin are comparable and carry the weak CP phases, i.e., $\gamma$ and $\beta$, from Fig.~\ref{fig:BRACPpi}(b) the CPA of $B^-_c \to \bar D^0 \pi^-$ can be of $O(1)$.  

\begin{figure}[phtb]
\includegraphics[scale=0.5]{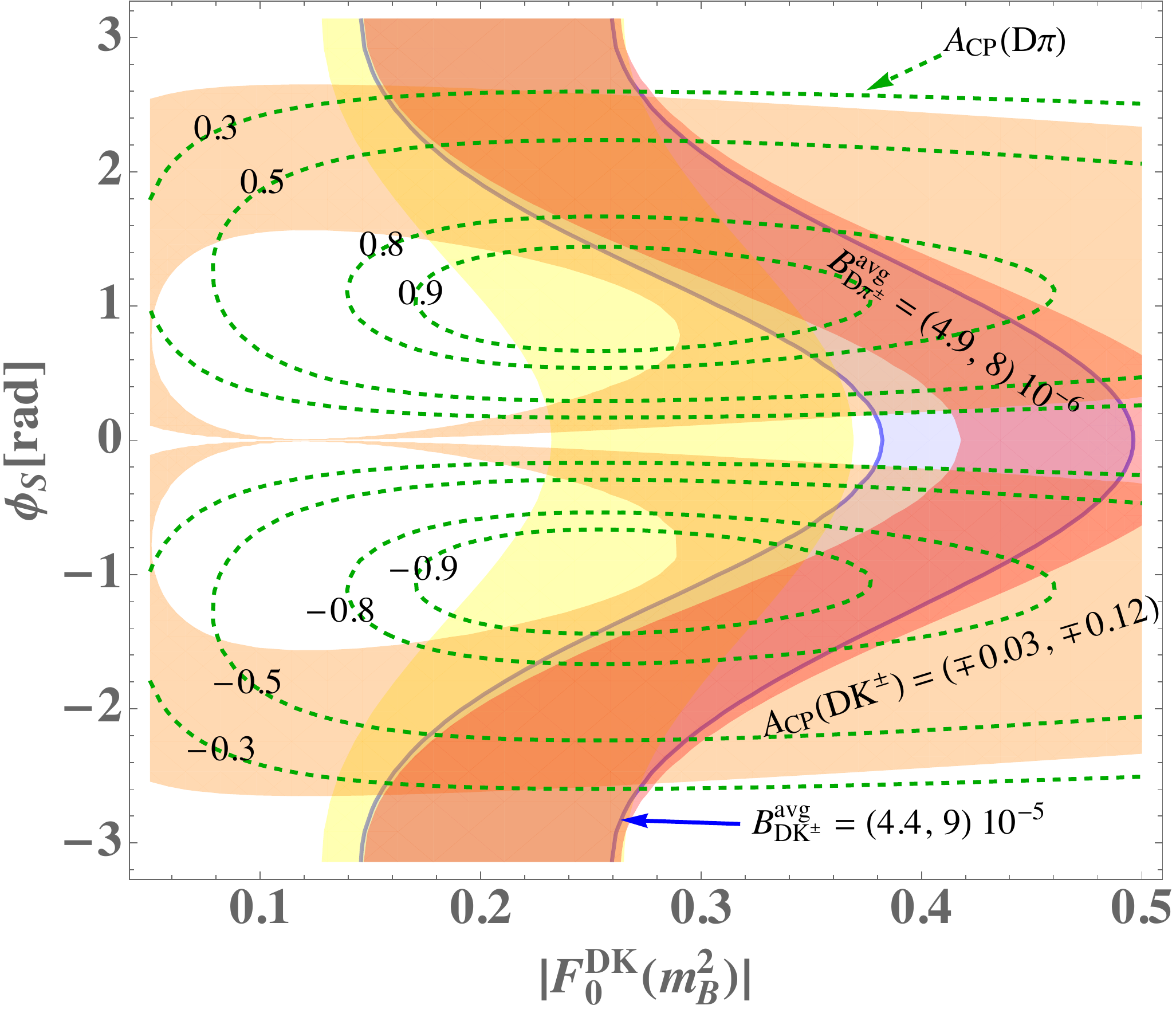}

 \caption{ ${\cal B}(B^-_u \to D^- \bar K^0)=(1,3.1)\times 10^{-7}$ (yellow), ${\cal B}^{\rm avg}_{D^0K^\pm}=(4.4,9)\times 10^{-5}$ (blue), $A_{CP}(B^\pm_c \to D^0 K^\pm)=(\mp 0.03, \mp 0.12)$ (orange), ${\cal B}^{\rm avg}_{D^0 \pi^\pm}=(4.9, 8)\times 10^{-6}$ (red), and $A_{CP}(B^\pm_c \to D^0 \pi^\pm)=( \pm 0.1, \pm 0.9)$ (dashed) in $\phi_S-|F^{DK}_0|$ plane.}
\label{fig:combine}
\end{figure}

\begin{figure}[phtb]
\includegraphics[scale=0.45]{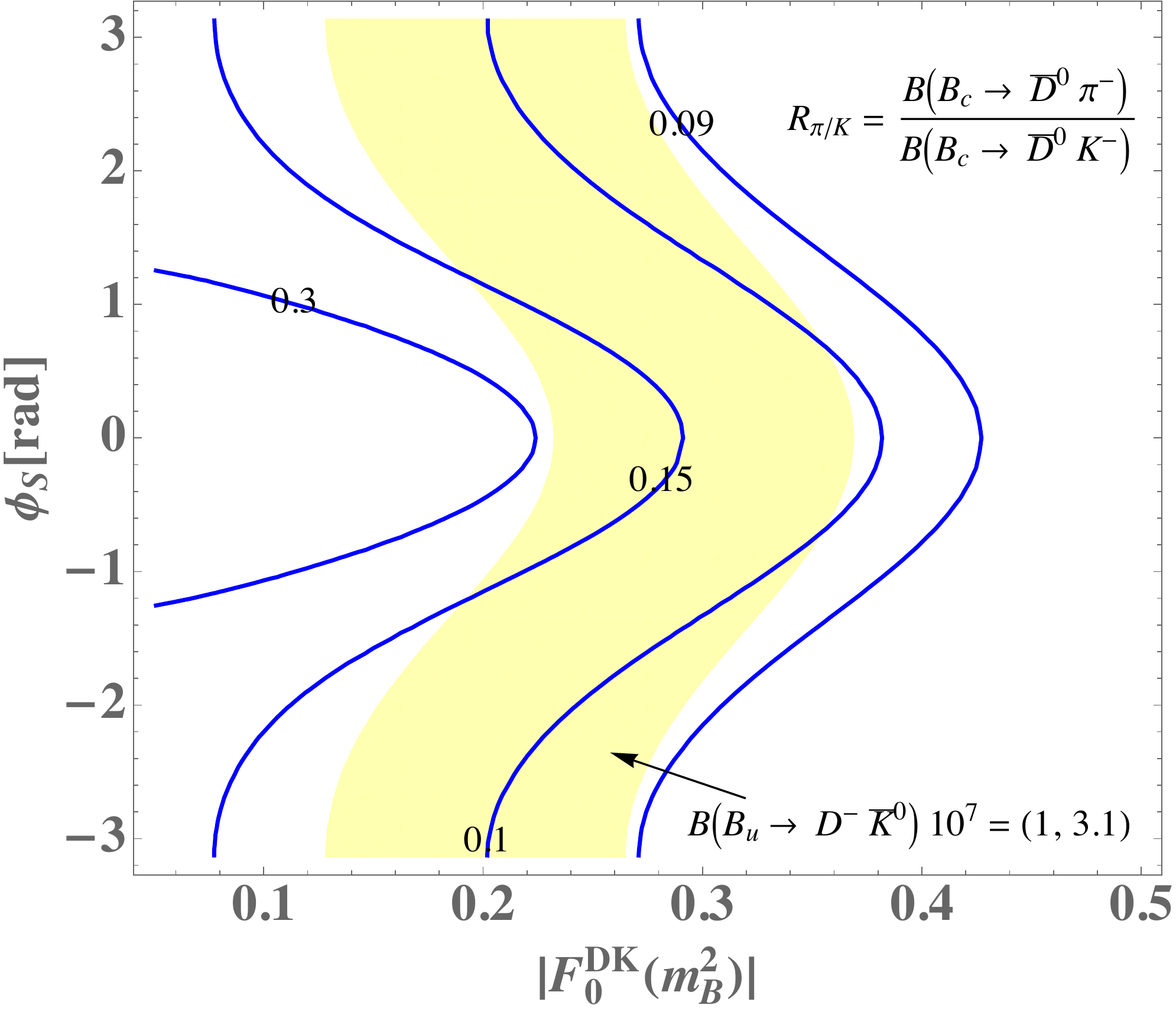}
 \caption{Ratio of the branching fraction of $B^-_c \to \bar D^0 \pi^-$ to $B^-_c \to \bar D^0 K^-$, where the shaded area denotes the $1<{\cal B}(B^-_u \to D^- \bar K^0)10^{7}<3.1$. }
\label{fig:Rpi_K-d}
\end{figure}

We show the BRs for $B^\pm_c \to  D^0 \pi^\pm$ and the CPA with some selected values of the strong phase in Table~\ref{tab:BRBcDpi}, where $f^{B_cD}_0=0.20$ and $|F^{DK}_0|=0.24$ are used. For clarity, we show the ranges ${\cal B}(B^-_u \to D^- \bar K^0)=(1,3.1)\times 10^{-7}$ (yellow), ${\cal B}^{\rm avg}_{D^0K^\pm}=(4.4,9)\times 10^{-5}$ (blue), $A_{CP}(B^\pm_c \to D^0 K^\pm)=(\mp 0.03, \mp 0.12)$ (orange), ${\cal B}^{\rm avg}_{D^0 \pi^\pm}=(4.9, 8)\times 10^{-6}$ (red), and $A_{CP}(B^\pm_c \to D^0 \pi^\pm)=( \pm 0.1, \pm 0.9)$ (dashed) as a function of $\phi_S$ and $|F^{DK}_0|$ in Fig.~\ref{fig:combine}. Due to $V_{cd}<0$, the CPAs for the $D^0 K^\pm$ and $D^0\pi^\pm$ modes are opposite in sign. Since we cannot precisely  determine the strong phase $\phi_S$, the allowed CPA for $D^0 \pi^-$ mode in our analysis is wide. In addition, the  ratio of branching fraction of $B^-_c \to \bar D^0 \pi^-$ to $B^-_c \to \bar D^0 K^-$, denoted by $R_{\pi/K}$ is shown in Fig.~(\ref{fig:Rpi_K-d}). The range of the ratio can be $0.1-0.2$ when ${\cal B}(B^-_u \to D^- \bar K^0)=(1,3.1)\times 10^{-7}$.

\begin{table}[htp]
\caption{ Branching ratios for $B^-_c \to \bar D^0  \pi^-$ and $B^+_c \to D^0 \pi^+$ and CP asymmetry with benchmarks of  $\phi_S$, where $f^{B_cD}_0=0.20$ and $|F^{DK}_0|=0.24$ are used.  }
\begin{tabular}{c|ccccc} \hline\hline
$\phi_S$ & ~~~$0$~~~~& ~~~~$+\pi/2$& ~~~$+2\pi/3$~~~& ~~~$-\pi/2$~~~ &~~~ $-2\pi/3$~~~\\ \hline
${\cal B}_{\bar D^0 \pi^-}10^{6}$ &   2.21  & 8.76 & 9.49 &  0.74   & 2.55 \\ \hline
${\cal B}_{D^0 \pi^+}10^{6}$  & 2.21 & 0.74 & 2.55 & 8.76 & 9.49  \\ \hline
${\cal B}^{\rm avg}_{D \pi^\pm}10^{6}$  & 2.21 & 4.75 & 6.02 & 4.75 & 6.02  \\ \hline
$A_{CP}$ &   0 & 0.84 & 0.58 & -0.84 & -0.58 
 \\ \hline \hline
\end{tabular}
\label{tab:BRBcDpi}
\end{table}%

\begin{figure}[phtb]
\includegraphics[scale=0.35]{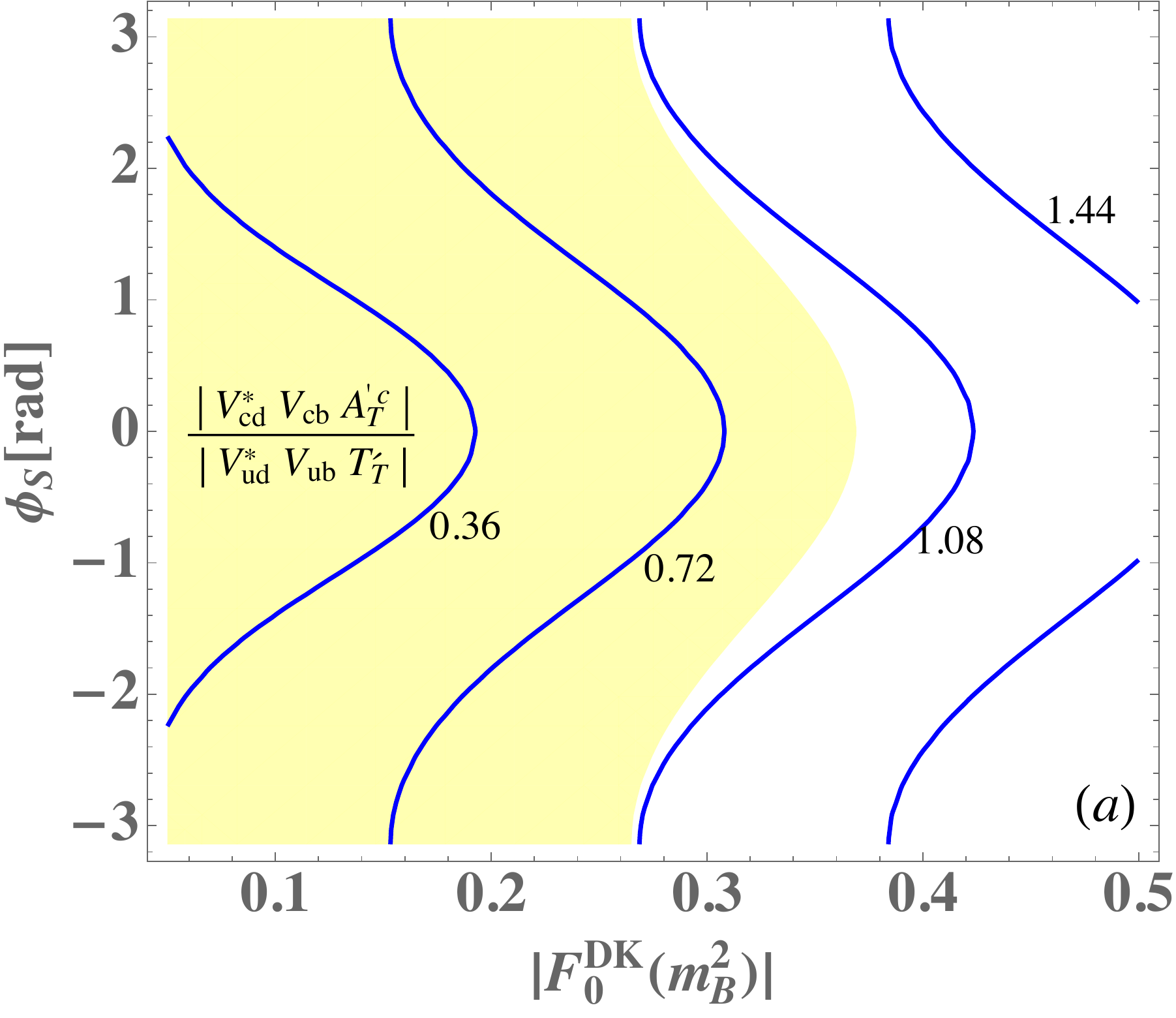}
\includegraphics[scale=0.35]{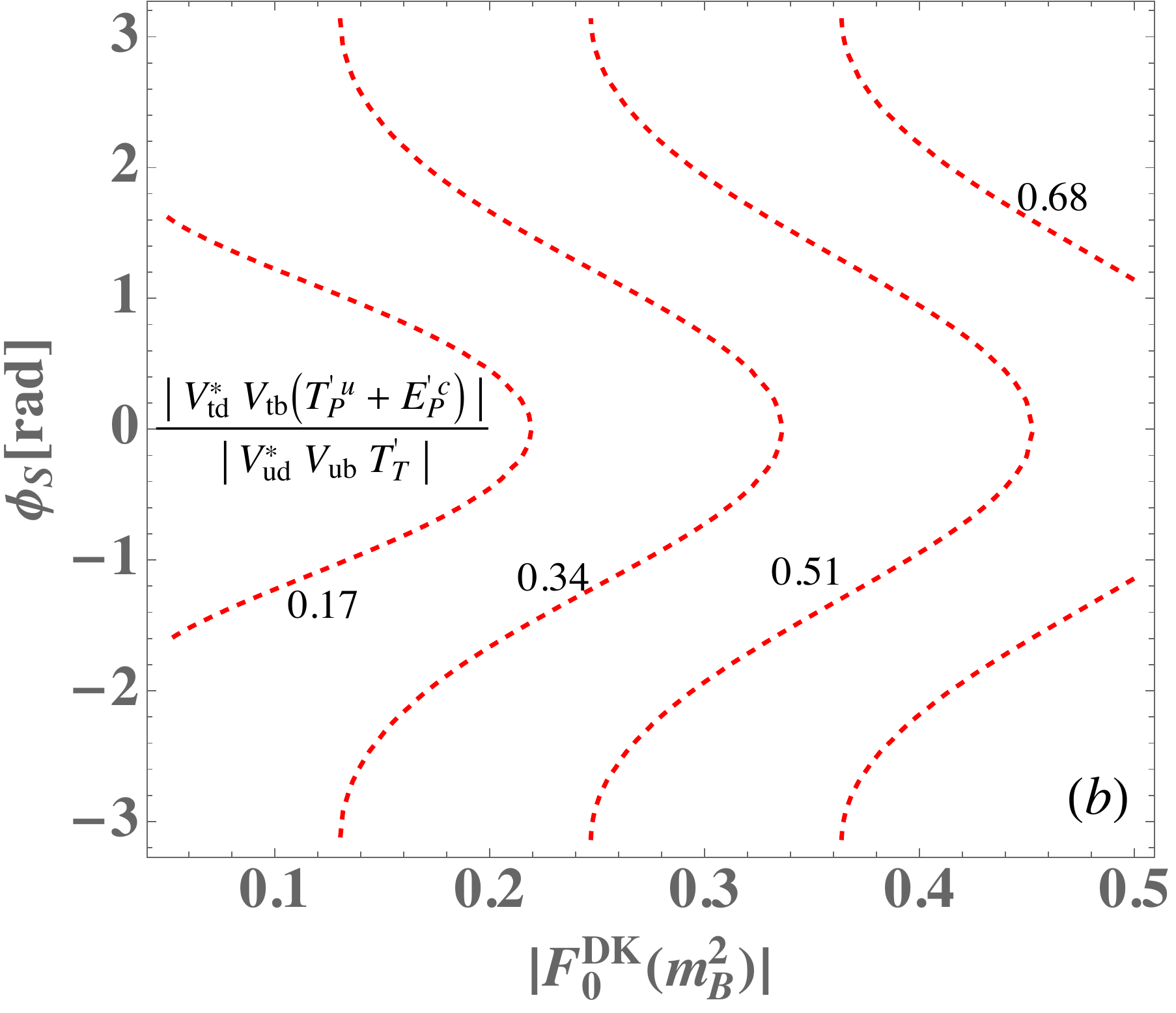}
\includegraphics[scale=0.35]{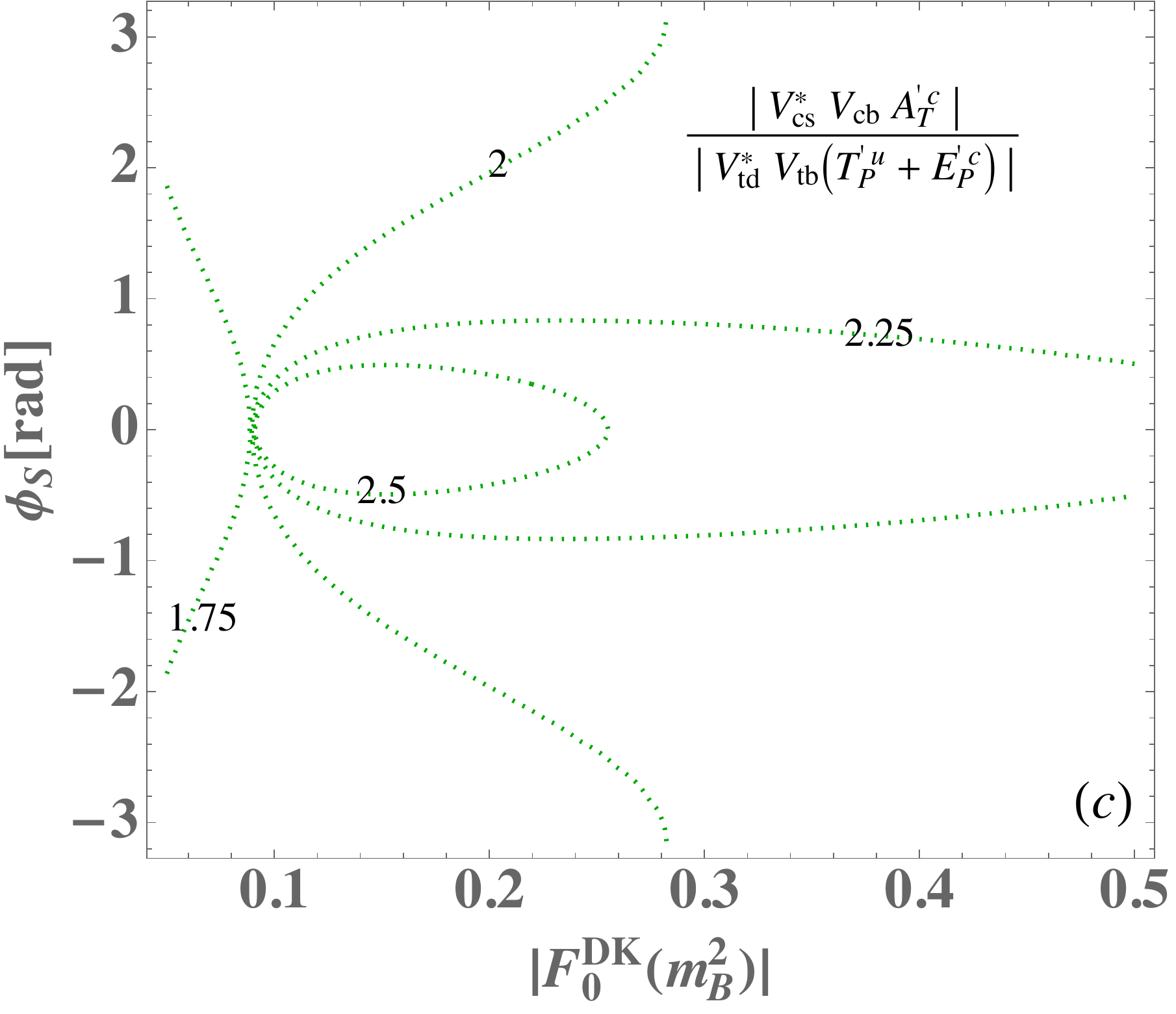}
\includegraphics[scale=0.35]{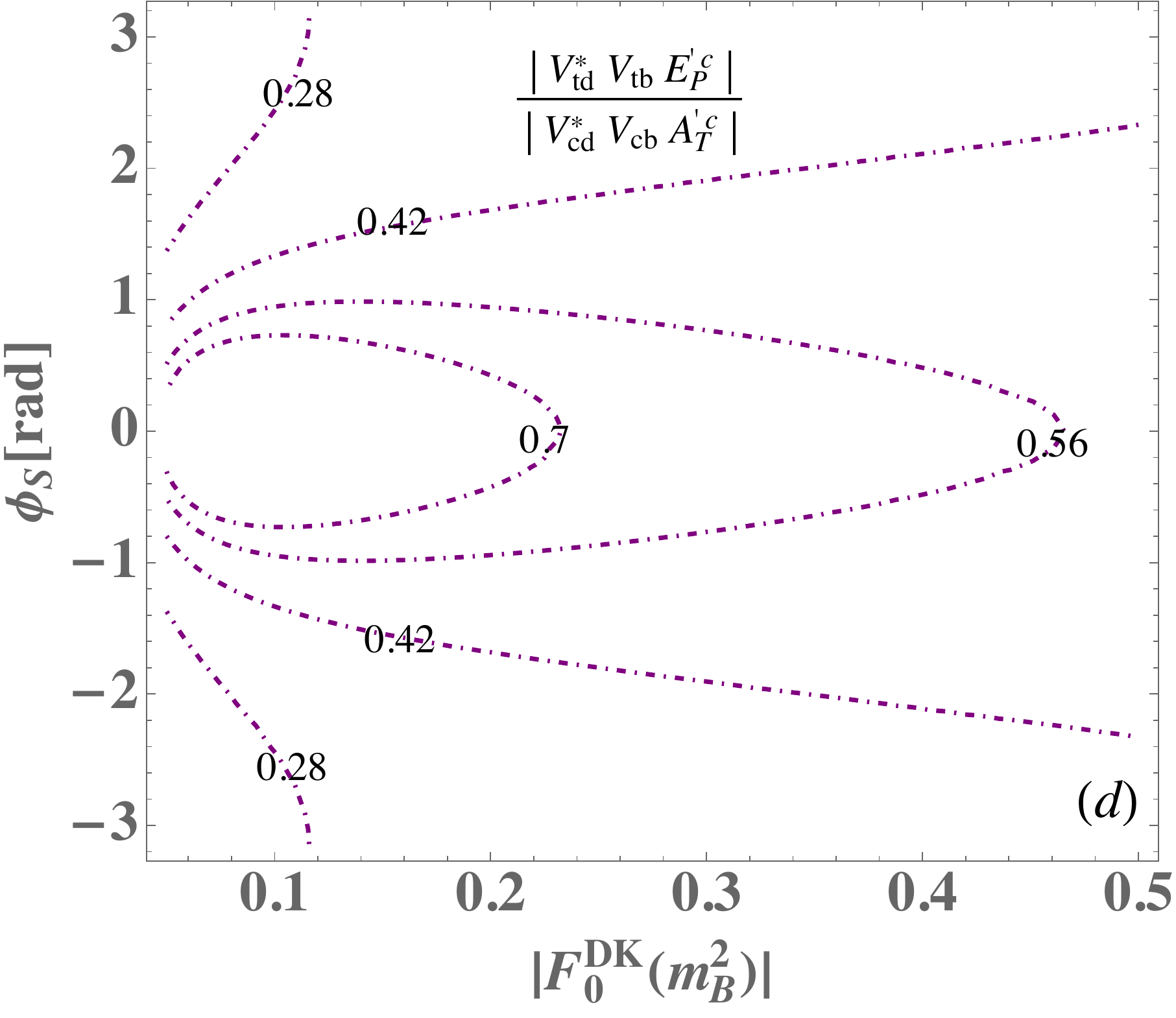}
 \caption{Contours for (a) $|V^*_{cd} V_{cb} A'^c_T|/|V^*_{ud} V_{ub} T'_T|$, (b) $|V^*_{td} V_{tb} (T'^u_P+E'^c_P)|/|V^*_{ud} V_{ub} T'_T|$, (c) $|V^*_{cd} V_{cb} A'^c_T|/|V^*_{td} V_{tb} (T'^u_P+E'^c_P)|$, and (d) $|V^*_{td} V_{tb} E'^c_P|/|V^*_{cd} V_{cb} A'^c_T|$ as a function of $\phi_S$ and $|F^{DK}_0(m^2_B)|$, where the shaded area in (a) is the bound of ${\cal B}(B_c \to D^- K^0)<3.1 \times 10^{-7}$, and  $f^{B_cD}_0=0.20$ is taken. }
\label{fig:Rpia-d}
\end{figure}

In order to understand the contribution of each component in the decay amplitude of Eq.~(\ref{eq:BcDpi}), we  present the ratios of $|V^*_{cd} V_{cb} A'^c_T|/|V^*_{ud} V_{ub} T'_T|$, $|V^*_{td} V_{tb} (T'^u_P+E'^c_P)|/|V^*_{ud} V_{ub} T'_T|$, $|V^*_{cd} V_{cb} A'^c_T|/|V^*_{td} V_{tb} (T'^u_P+E'^c_P)|$, and $|V^*_{td} V_{tb} E'^c_P|/|V^*_{cd} V_{cb} A'^c_T|$ in Fig.~\ref{fig:Rpia-d}(a)-(d), where $f^{B_cD}_0=0.20$ is used. The tree-transition dominance in $B^-_c \to \bar D^0 \pi^-$ can be verified  from plots (a) and (b); and in some regions, the $A'^c_T$ effect can be comparable to the $T'_T$. When $V_{cs}$ and $V_{ts}$ in $\bar D^0 K^-$ mode are replaced with $V_{cd}$ and $V_{td}$ in $\bar D^0 \pi$, the tree-annihilation and penguin contributions are roughly multiplied by a factor of $\lambda$ at the same time; therefore,  the ratios of tree-annihilation to penguin and penguin-annihilation to tree-annihilation are similar to the cases in $B^-_c \to \bar D^0 K^-$. Hence, although $T'_T$ topology dominates in $\bar D^0 \pi^-$ channel, the $A^c_T$ contribution is also important.

\section{Summary}

We  studied the $B^-_c \to (\bar D^0 K^-, \bar D^0 \pi^-)$ decays  using a phenomenological analysis, where we employed the $B\to KK$, $B^-_u\to D^- \bar K^0$, and $\bar B_d \to D_s K^-$ decays to determine the nonfactorization effect  and to limit the factorization effect of an annihilation process. According to our study, a factorizable tree-annihilation should dominate the decay amplitude of $B^-_c \to \bar D^0 K^-$. The relative magnitude of $T^u_P$ and $E^c_P$ depends on the time-like form factor $F^{DK}_0$; that is,  when $F^{DK}_0 \lesssim 0.2$, $|T^u_P|> |E^c_P|$. If we take the  branching ratio of $B^-_u \to D^- \bar K^0$ to be $(1-3.3) \times 10^{-7}$,  the branching ratio of $B^-_c \to \bar D^0 K^-$ is obtained in the range $(4.4-9) \times 10^{-5}$, where the result falls within $1.5\sigma$ of $(10.01\pm 3.40)\times 10^{-5}$, which is extracted from LHCb result with ${\cal B}(B^-_c \to J/\Psi \pi^-)=(7.7\pm 1.1) \times 10^{-4}$.  The CP asymmetry of $B^-_c \to \bar D^0 K^-$ is derived from the interferences between the small tree-transition and tree-annihilation and from those between the small tree-transition and penguin-annihilation;  as a result, the magnitude of the CP asymmetry is less than approximately $10\%$.

$B^-_c \to \bar D^0 \pi^-$ should be dominated by the tree-transition contribution due to the CKM factor $V^*_{ud} V_{ub}$ and Wilson coefficient $a_1$. Although the CKM factor $V^*_{ud} V_{cb}$ in the tree-annihilation has an extra Wolfenstein parameter suppression, due to $|V_{ub}| < |V^*_{cd} V_{cb}|$, the tree-annihilation topology still play an important role  in $B^-_c \to \bar D^0 \pi^-$. When we take ${\cal B}(B^-_c \to \bar D^0 K^-)\approx (4.4-9) \times 10^{-5}$, the corresponding BR for $B^-_c \to \bar D^0 \pi$ is ${\cal B}(B^-_c \to \bar D^0 K^-)\approx (4.9-8) \times 10^{-6}$. Due to the contributions from  the tree and penguin being comparable, the CP asymmetry of $B^-_c \to \bar D^0 \pi^-$, which arise from the interferences between the tree-transition and the penguin, between the tree-annihilation and penguin, and between the tree-transition and tree-annihilation,  can be of the order of one. 

 In this study, we also predict ${\cal B}(B^-_u \to D^- \bar K^0)<3.1 \times 10^{-7}$, ${\cal B}(B^-_c \to D^- \bar K^0) \approx {\cal B}(B^-_c \to \bar D^0 K^-)$, ${\cal B}(B^-_c \to K^- K^0)\approx (6.99 \pm 1.34)\times 10^{-7}$, and ${\cal B}(B^-_c \to J/\Psi \pi^-) \approx (7.7 \pm 1.1)\times 10^{-4}$.

\section*{Acknowledgements}

This work was partially supported by the Ministry of 
Science and Technology of Taiwan,  
under grants MOST-106-2112-M-006-010-MY2 (CHC) and MOST-106-2811-M-006-041(YHL).

%%%%%%%%%%%%%%%%%%%%%%%%%%%%%%%%%%%%%%%%%%%%%%%%%%%%%%

\end{document}